\documentclass[prx,aps,reprint,noshowpacs,superscriptaddress,floatfix,letterpaper,longbibliography]{revtex4-2}
\usepackage{amsmath,amssymb,amsbsy,amsfonts,amsthm,bbm,bm,mathtools,mathrsfs,bbold}
\usepackage{color}
\usepackage{physics}
\usepackage{xfrac}
\usepackage[dvipsnames]{xcolor}
\usepackage[colorlinks=true,citecolor=MidnightBlue,linkcolor=MidnightBlue,urlcolor=MidnightBlue]{hyperref}
\usepackage{empheq,ragged2e}
\usepackage{pgfplots}
\usepackage{stackengine}
\usepackage{relsize}
\usepackage[inline]{enumitem}
\usepackage[normalem]{ulem}
\usepackage{comment}
\usepackage{import}

\usepackage{pgfplots}
\pgfplotsset{compat = newest}
\usetikzlibrary{arrows,intersections}
\usepackage{tikz-3dplot}
\usepackage{tikz}

\usepackage[mathscr]{euscript}
\usepackage{calligra}
\DeclareMathAlphabet{\mathcalligra}{T1}{calligra}{m}{n}
\DeclareFontShape{T1}{calligra}{m}{n}{<->s*[2.2]callig15}{}

\newcommand{\stability}{{\bm{A}}}
\newcommand{\jacobian}{{\bm{M}}}
\newcommand{\ex}{{\bm{x}}}
\newcommand{\bphi}{{\bm{\phi}}}

\newcommand{\bpsi}{{\bm{\psi}}}
\newcommand{\bPsi}{{\bm{\Psi}}}
\newcommand{\yu}{{\bm{u}}}
\newcommand{\brho}{{\bm{\varrho}}}
\newcommand{\bxi}{{\bm{\xi}}}
\newcommand{\stabilityH}{{\bm{\mathcal{H}}}}
\newcommand{\Hamiltonian}{H}
\newcommand{\bmP}{{\bm{p}}}
\newcommand{\bmQ}{{\bm{q}}}
\graphicspath{{../plots/}} 

\usepackage{tcolorbox}
\usepackage{multirow}
\definecolor{new_blue}{RGB}{9, 136, 232}
\newtcolorbox{mybox}[3][]
{%
	colframe = #2!25,
	colback  = #2!10,
	coltitle = #2!20!black,  
	title    = {#3},
	#1,
}
\tcbuselibrary{breakable}

\usepackage{pgfplots}
\pgfplotsset{compat = newest}
\usetikzlibrary{arrows,intersections}
\usepackage{tikz-3dplot}
\usepackage{tikz}

\begin{document}

\def\xlist{4}
\def\ylist{4}

\title{Density matrix formulation of dynamical systems}

\author{Swetamber~Das}
\author{Jason~R.~Green}
\email[]{jason.green@umb.edu}
\affiliation{Department of Chemistry,\
	University of Massachusetts Boston,\
	Boston, MA 02125
}
\affiliation{Department of Physics,\
	University of Massachusetts Boston,\
	Boston, MA 02125
}

\date{\today}

\begin{abstract}

Physical systems that dissipate, mix and develop turbulence also irreversibly transport statistical density.
In statistical physics, laws for these processes have a mathematical form and tractability that depends on whether the description is classical or quantum mechanical.
Here, we establish a theory for density transport in any classical dynamical system that is analogous to the density matrix formulation of quantum mechanics.
Defining states in terms of a classical density matrix leads to generalizations of Liouville's theorem and Liouville's equation, establishing an alternative computationally-tractable basis for nonequilibrium statistical mechanics.
The formalism is complete with classical commutators and anti-commutators that embed measures of local instability and chaos and are directly related to Poisson brackets when the dynamics are Hamiltonian.
It also recovers the traditional Liouville equation and the Liouville theorem by imposing trace preservation or Hamiltonian dynamics.
Applying to systems that are driven, transient, dissipative, regular, and chaotic, this formalism has the potential for broad applications.

\end{abstract}

\maketitle
\section{Introduction}

Whether classical and quantum mechanical, the transport of statistical density is our primary means of making predictions of macroscopic, nonequilibrium behavior from microscopic dynamics~\cite{zwanzig2001nonequilibrium}.
Classically, Jacobi's form of Liouville's equation of motion for the phase space density of mechanical systems is the foundation of classical statistical mechanics~\cite{tolman1979principles}.
Its usefulness derives largely from its many forms and approximations, including the Boltzmann equation, the Vlasov approximation, the Bogoliubov-Born-Green-Kirkwood-Yvon hierarchy, that underlie applications across physics and chemistry~\cite{zwanzig2001nonequilibrium}.
In quantum mechanics, the Liouville-von Neumann equation describes the evolution of the density operator~\cite{Neu1927}; it is the fundamental equation of quantum statistical mechanics and a main ingredient in quantum computing, tomography, and decoherence~\cite{Blu2012}.
To translate between the classical and quantum mechanical Liouville equations, one can use Dirac's rule~\cite{dirac1981principles} of replacing Poisson brackets by commutators.
Here, we establish a density matrix formalism for classical systems, with not only a density matrix but also a classical commutator, supplanting Dirac's heuristic with a more direct correspondence between these physical theories.

Operator-theoretic methods, such as Frobenius-Perron and its dual Koopman formalism~\cite{Koopman315}, give a formal analogy to quantum mechanics by lifting the description of classical systems to infinite dimensions~\cite{neumann_zur_1932,*neumann_zusatze_1932}.
They preserve global nonlinear features and guarantee exact linearization of the dynamics, providing useful connections between classical dynamical systems and statistical physics~\cite{dorfman1999introduction,gaspard2005chaos}.
However, they can be difficult to apply to systems under active external control and to find observables representing the nonlinear system in the lifted linear space~\cite{lusch_deep_2018}.
Symmetries can make the calculation of the Koopman operator approximation and its spectral properties more efficient~\cite{salova_koopman_2019}, but, in practice~\cite{rowley2009,budisic_applied_2012}, the number of variables must be truncated to finite-dimensions (e.g., through extended~\cite{williams_datadriven_2015,korda_convergence_2018} or kernel~\cite{williamsrowly2015} dynamic mode decomposition~\cite{schmid_2010}).
While reminiscent of quantum mechanics, these are classical theories that add weight to the question of whether other formulations of quantum mechanics might have classical counterparts that are fundamental to statistical physics~\cite{BudRoh2017}.

While Liouville's equation is the foundation of nonequilibrium statistical physics, many theories avoid, approximate, or subject Liouville's equation to model specific solutions~\cite{zwanzig2001nonequilibrium}.
Here, we construct a classical density matrix formulation of dynamical systems on the local stability of nonlinear dynamics~\cite{Eckmann1985Ergodic} -- Lyapunov exponents and vectors~\cite{PikovskyP16}.
The infinitesimal perturbations, Lyapunov vectors, we use to define the density matrix have been used to analyze rare trajectories~\cite{TailleurK07}, jamming~\cite{banigan_chaotic_2013}, nonequilibrium self-assembly~\cite{GreenCGS13}, equilibrium and nonequilibrium fluids~\cite{evans1990statistical, Bosetti_2014,das_self-averaging_2017}, and critical phenomena~\cite{dascritical2019}.
From these finite-dimensional vectors, we derive a classical analogue of the von Neumann equation for the density matrix dynamics.
We show this classical density matrix is similar to the dual metric tensor and that its determinant evolves according to a generalized Liouville equation and satisfies a generalized Liouville theorem.
And, by imposing a norm-preserving dynamics with Lyapunov exponents not only normalizes the density matrix, it reinstates the form of the usual Liouville equation for generic, non-Hamiltonian dynamical systems.
All together, these results establish a computationally-tractable basis for nonequilibrium statistical mechanics grounded in dynamical systems theory.

The structure and this formalism suggests the possibility of other classical counterparts to elements of quantum theory, including uncertainty relations and speed limits; In Ref.~\cite{das2021speed}, we use the classical density matrix theory here to derive a family of time-information uncertainty relations that set speed limits on the evolution of (observables of) arbitrary dynamical systems.

\begin{figure*}[!htb]
\vspace{-2cm}
\includegraphics[width=0.49\linewidth, height=0.3\textheight]{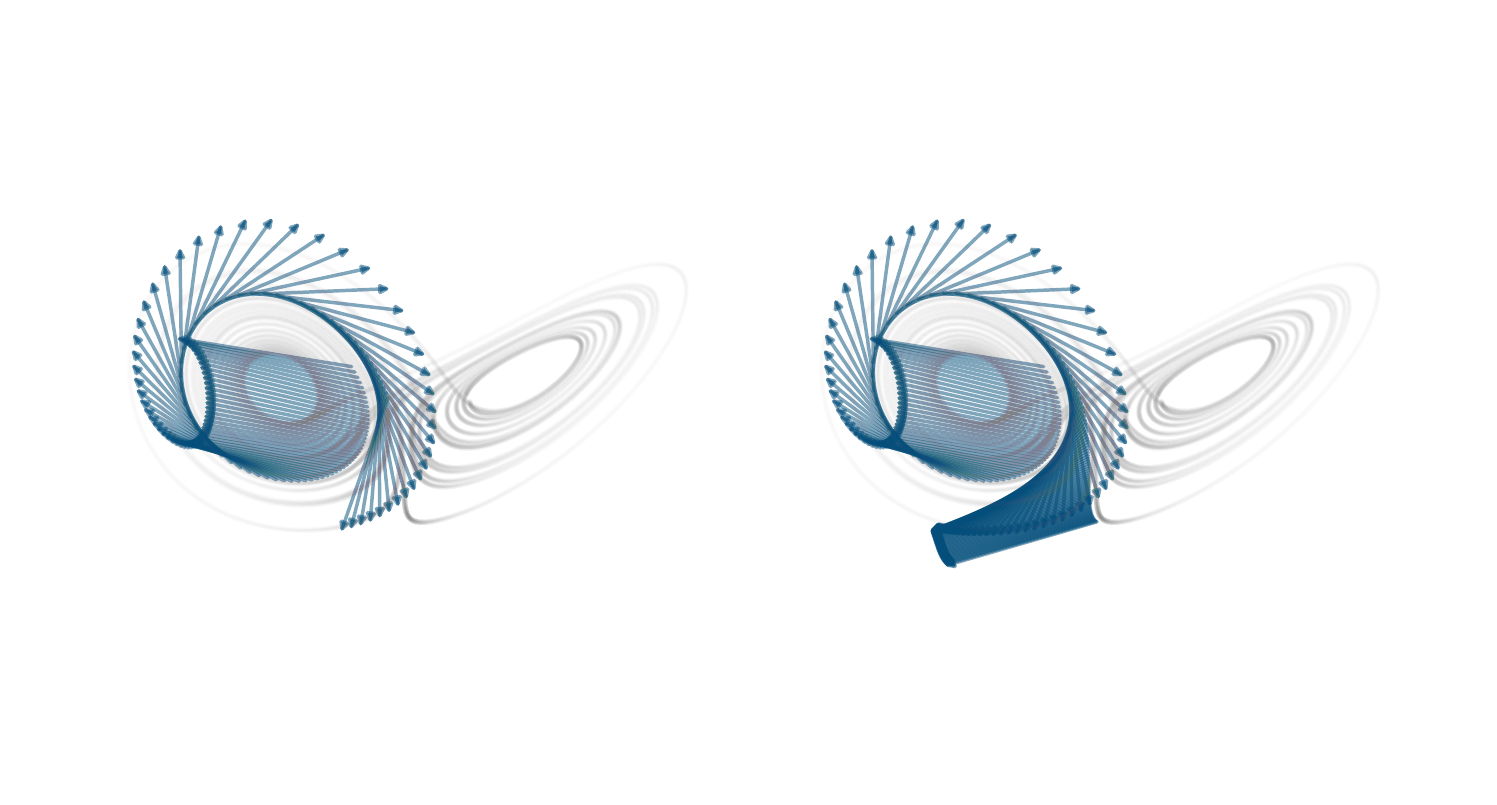}\hspace{-0.5cm}
\includegraphics[width=0.49\linewidth, height=0.3\textheight]{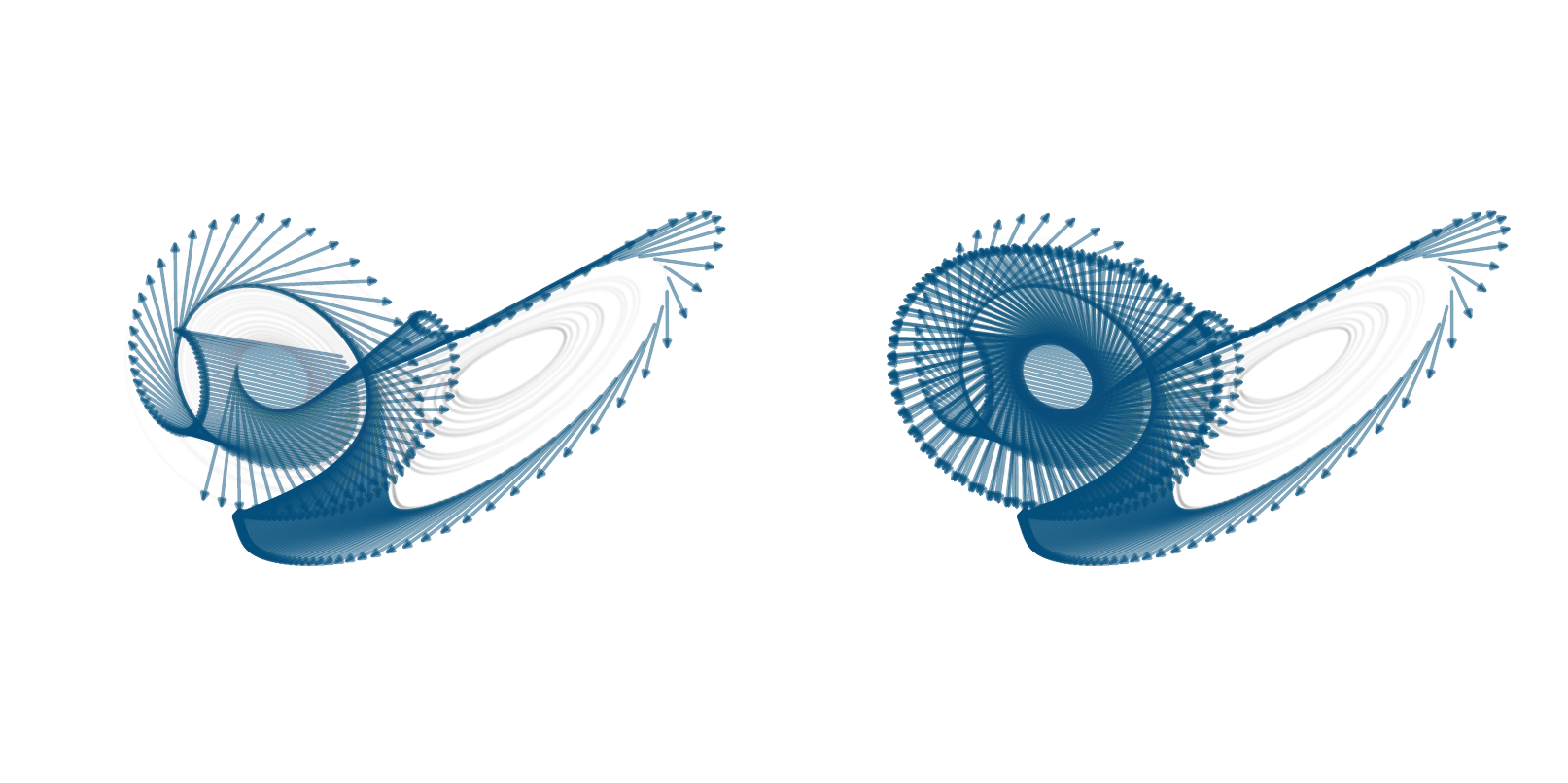}
\vspace*{-2cm}
\caption{\label{fig:Lorenz_attractors}Snapshots of the strange attractor and an initially random, unit perturbation transported along a chaotic solution of the Lorenz system. Parameters are those originally used by Lorenz~\cite{Lorenz63}: $\sigma = 10$, $\beta = 8/3$, $\rho = 28$.}

\end{figure*}

To start, we define an unnormalized density matrix from the linearization of the classical dynamics, Sec.~\ref{sec:dynamics_unnorm}.
The properties of this density matrix lead to a generalization of Liouville's theorem and equation, Sec.~\ref{sec:GLE/GLT}.
For Hamiltonian dynamics, Sec.~\ref{sec:PB}, we show the reduction to the usual Liouville theorem and equation and establish a connection to Poisson brackets.
The dynamics of a normalized density matrix, Sec.~\ref{sec:dynamics_norm}, transform the generalized Liouville's theorem and equation to the usual form, Sec.~\ref{sec:GLE_norm}.
The basis chosen determines the representation of the density matrix, Sec.~\ref{sec:basis}, which is a common consideration in dynamical systems theory.
This formalism is illustrated analytically for the linear and damped harmonic oscillator in Sec.~\ref{sec:case_studies}.

\section{Dynamics of the classical unnormalized density matrix}\label{sec:dynamics_unnorm}
\noindent Consider a generic dynamical system with state-space variables $\{x^i\}$.
At any moment in time, these variables together mark a point $\ex(t) := [x^1(t),x^2(t),\ldots,x^n(t)]^\top$ in an $n$-dimensional state space $\mathcal{M}$ that evolves according to: $\dot{\ex} = \boldsymbol{F}[\ex(t)]$.
Perturbations to the system will also evolve under the flow of the dynamics.
Because of their analytical and computational tractability, infinitesimal perturbations $\ket{\delta\ex(t)}:= [\delta x^1(t), \delta x^2(t), \ldots, \delta x^n(t)]^\top\in T\mathcal{M}$ and their linearized dynamics are a well established means of analyzing the stability of nonlinear dynamical systems~\cite{PikovskyP16}.
These perturbations represent an uncertainty about the initial condition and stretch, contract, and rotate over time,
\begin{equation}
  \ket{\delta\dot{\ex}(t)} = \stability[\ex (t)]\ket{\delta \ex(t)},
  \label{equ:EOM_per_ket2}
\end{equation}
with the phase point. Their time evolution is governed by the local stability matrix $\stability:=\stability[\ex(t)] = \grad\boldsymbol{F}$ with elements
$(\stability)^i_{j}=\partial \dot{x}^i(t)/\partial x^j(t)$.
Figure~\ref{fig:Lorenz_attractors} shows a unit perturbation vector as it is transported across the Lorenz attractor.
Surrounding the phase point $\ex(t)$ is an infinitesimal $k$-dimensional phase space volume that transforms its shape over time as it evolves.
The volume we take to be spanned by a finite set $\{\ket{\delta\bpsi_i}\}$ of $k\leq n$ basis vectors $\ket{\delta\bpsi_i} \in T\mathcal{M}$ ($i=1,2,\ldots, k$) that also obey the linearized dynamics (e.g., Lyapunov vectors~\cite{f_ginelli_characterizing_2007,Wolfe2007}).
If the dynamics are Hamiltonian, then according to Liouville's theorem, the volume spanned by these tangent states is conserved at all times.

Comparing to quantum mechanics, the evolution of classical, finite-dimensional, real tangent space vectors according to Eq.~\ref{equ:EOM_per_ket2} is analogous to the evolution of infinite-dimensional, complex Hilbert space vectors according to Schr\"odinger's equation~\cite{dirac1981principles}.
An alternative representation of quantum states, however, is the density operator formalism, which provides a basis for quantum technology, dissipative decoherence, and statistical mechanics~\cite{Fan1957,Haa1961,Blu2012}.
Because of the power of this formulation of quantum mechanics, we break from traditional classical dynamical systems, constructing this formulation by defining the classical state with a density operator:
\begin{equation}
  \bxi(t) := \sum_{i=1}^k\dyad{\delta \bpsi_i(t)}.
\end{equation}
This \textit{unnormalized} matrix form of this operator represents an alternative state of a classical dynamical system. 
Expressed here in the $\{\delta\bpsi_i\}$ basis, it is the outer product of perturbation vectors or what Gibbs called the dyadic~\cite{gibbs1901vector} product.

Partitioning the stability matrix $\stability = \stability_+ + \stability_-$ into its symmetric and anti-symmetric parts, $\stability_\pm = \frac{1}{2}(\stability \pm \stability^\top)$, the time evolution of $\bxi$ is:
\begin{equation}\label{eq:EOM_xi}
  \frac{d\bxi}{dt} = \{\stability_+,\bxi\}+[\stability_-,\bxi].
\end{equation}
Its solution,
\begin{align}\label{eq:op_M}
  \bxi(t) = \jacobian(t,t_0)\bxi(t_0)\jacobian^\top(t,t_0),
\end{align}
is in terms of the non-normal propagator, $(\jacobian)^i_{j}=\partial x^i(t)/\partial x^j(t_0)$ (Appendix~\ref{sec:EOM_LV}) when $\bxi$ is built from the tangent vectors $\{\ket{\delta \ex^i}\}$ evolved by $\jacobian$.

The generally non-symmetric stability matrix $\stability$ plays the role of the quantum mechanical Hamiltonian in the \textit{classical} commutator $[\boldsymbol{X},\boldsymbol{Y}]=\boldsymbol{Y}\boldsymbol{X}-\boldsymbol{X}\boldsymbol{Y}$ and anti-commutator $\{\boldsymbol{X},\boldsymbol{Y}\}=\boldsymbol{Y}\boldsymbol{X}+\boldsymbol{X}\boldsymbol{Y}$.
Two facts are striking about this solution: it is entirely computable from standard methods in dynamical systems theory~\cite{PikovskyP16} and it is a purely classical analogue of the Liouville-von Neumann equation in quantum dynamics.
Unlike its quantum counterpart, the determinant of $\bxi$ is directly related to Liouville's equation and theorem with this equation of motion, as we show next.

\subsection{Generalized Liouville's theorem/equation}\label{sec:GLE/GLT}

\noindent Liouville's theorem and equation are the foundation for nonequilibrium statistical mechanics~\cite{zwanzig2001nonequilibrium,dorfman1999introduction,gaspard2005chaos} and the point at which statistical mechanics departs from classical dynamics.
Here, we derive generalizations of both from the determinant of the unnormalized density matrix. To make this connection, consider the arbitrary set of basis vectors $\{\delta \bpsi_i\}$ that span the entire $n$-dimensional phase space volume, $d\mathcal{V}$, with the perturbation state $\bxi$. 
The (square of the) phase space volume is determined by the determinant $|\bxi|$,
which has the equation of motion (Appendix~\ref{sec:GLE}):
\begin{equation}\label{eq:gen_Liouville}
\frac{1}{2}\frac{d}{dt}\ln|\bxi(t)| = \grad\cdot \dot\ex = \Tr\stability_+ = \Lambda.
\end{equation}
Both this equation of motion and its solution,
\begin{equation}\label{eq:liouville_unnorm_den_mat}
  |\bxi(t)| = |\bxi(t_0)|\,e^{2\int_{t_0}^t\Lambda(t') \,dt'},
\end{equation}
depend on the divergence of the phase space velocity $\dot{\ex}$ or the phase space volume contraction/expansion rate $\Lambda = \Tr\stability_+ = \Tr\stability$.
This rate $\Lambda$ is (twice) the sum of the Lyapunov exponents~\cite{dorfman1999introduction}, which is related to physical quantities.
For example, the phase space contraction rate is related to the entropy production rate in fluid transport~\cite{dorfman1999introduction}.

\begin{figure}[t!]
	\includegraphics[width=1\columnwidth]{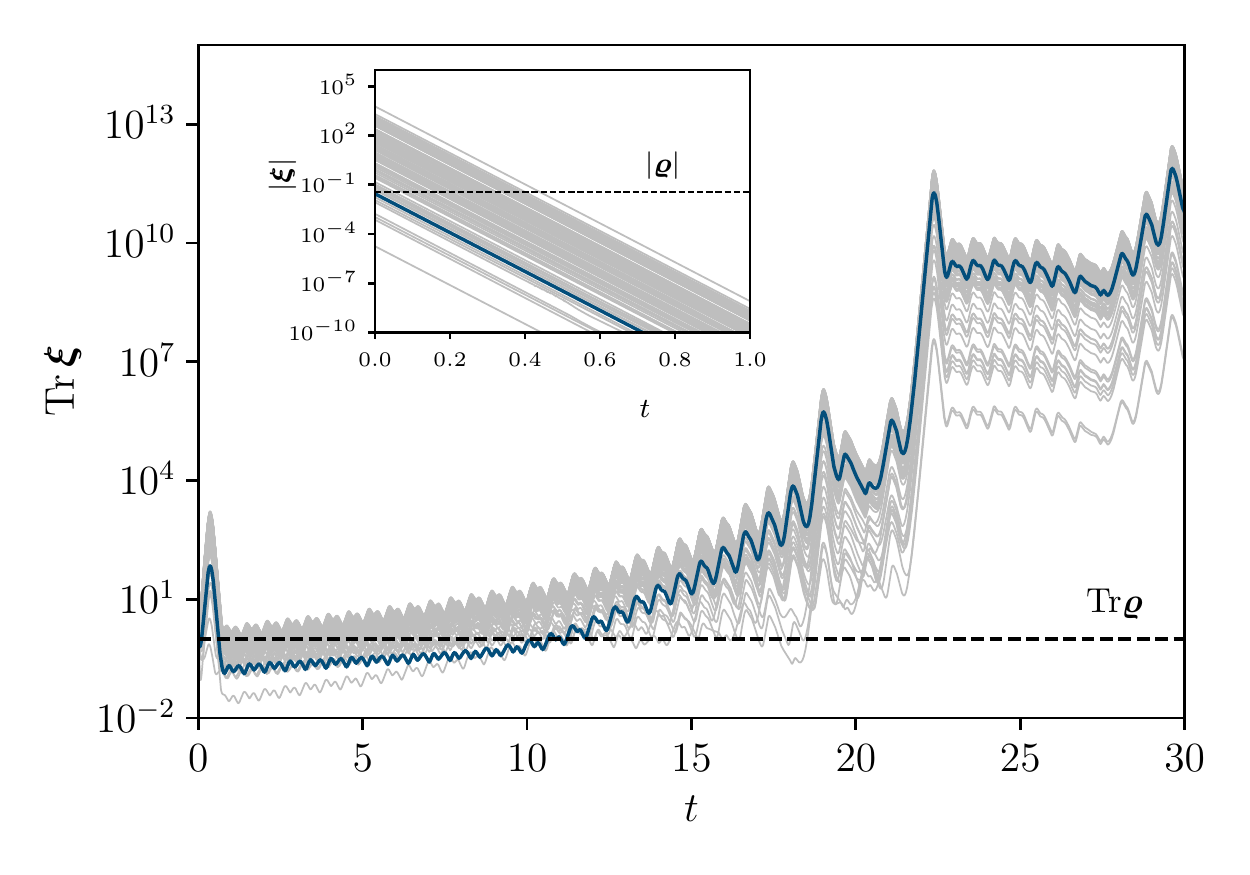}
	\caption{\label{fig:Lorenz_trace_det} The trace of the unnormalized density matrix, $\bxi(t)$, as a function of time for 100 random perturbations drawn from a uniform distribution. The inset shows the evolution of pure states $|\bxi(t)|$ for a set of 100 local perturbations. Data for the vector evolving on the Lorenz attractor in Fig.~\ref{fig:Lorenz_attractors} are shown in blue. The trace and determinant of the normalized density matrix $\brho$ (inset) are time-invariant (dashed).}
	\vspace{-.2in}
\end{figure}

From these results for the density matrix determinant, we can generalize Liouville's theorem and equation.
The geometric interpretation of Liouville's theorem is that the velocity field $\dot{\ex}$ has zero divergence: $\grad\cdot\dot{\ex} = \Tr(\stabilityH) = 0$,
and, consequently, the ``phase fluid'' flow is incompressible and phase space volumes are conserved $d\mathcal{V}(t)=d\mathcal{V}(t_0)$.
Here, $\stabilityH$ is the stability matrix for Hamiltonian dynamics with Hamiltonian $\Hamiltonian$. The phase space volume element $d\mathcal{V}(t)$ spanned by the basis vectors has a coordinate transformation: $d\mathcal{V}(t)=|\jacobian(t,t_0)|\,d\mathcal{V}(t_0)$. Combining this fact with the determinant of Eq.~\ref{eq:op_M}, we obtain a generalization of Liouville's theorem for general dynamical systems:
\begin{align}\label{eq:invariant_measure}
  |\bxi(t)|^{-\frac{1}{2}} d\mathcal{V}(t) &= |\bxi(t_0)|^{-\frac{1}{2}} d\mathcal{V}(t_0).
\end{align}
Any dynamics conserves the measure, $|\bxi|^{-\frac{1}{2}} d\mathcal{V}$ or $e^{-\int_{t_0}^t\Lambda(t')\,dt'}d\mathcal{V}$. 
For dissipative systems with $\Lambda < 0$, volumes contract at a rate $\Tr\stability$.
In the Lorenz model, for example, $\stability$ is constant, so $|\bxi|$ decays linearly on a semi-log scale with a slope proportional to $2\Tr\stability$ as shown in Fig.~\ref{fig:Lorenz_trace_det}.
But, when the dynamics are Hamiltonian, we recover the conventional form of Liouville's theorem~\cite{tolman1979principles} because $\Lambda = 0$ and $|\bxi|$ is a constant of motion.

Equation~\ref{eq:invariant_measure} also appears in Riemannian geometry as the transformation of a metric determinant on a Riemannian manifold of an arbitrary curvature endowed with a covariant metric tensor $\boldsymbol{g}_\bxi$.
Metrics have been considered previously in nonlinear dynamics~\cite{greeneIntroductionMetricTensor1989,thiffeaultCovariantTimeDerivatives2001}.
But, as we shown in Appendix~\ref{sec:GLE}, the (covariant) metric tensor is similar to the inverse of the unnormalized density matrix, $\bxi^{-1}$.
With this identification, a number of results follow.
Most immediate is that Eq.~\ref{eq:invariant_measure} becomes the transformation of the metric determinant: $\sqrt{g_\bxi(t)}d\mathcal{V}(t)=\sqrt{g_\bxi(t_0)}d\mathcal{V}(t_0)$ with $g_\bxi=|\bxi^{-1}|$.
However, what traditionally follows from Liouville's theorem is Liouville's equation, a formally exact equation of motion for the probability density in phase space.
This equation derives from another statement of Liouville's theorem: the density of representative points in the phase space is conserved along the trajectories of Hamiltonian systems, $d_t\rho(\ex)=0$~\cite{tolman1979principles}.
The density matrix also yields a generalization of this form of Liouville's equation (Appendix~\ref{sec:GLE}).

With the similarity of the unnormalized density matrix and the metric tensor, the flow compressibility accounts for the metric's compatibility with the dynamics~\cite{Ezra2004},
$-d_t\ln\sqrt{g_\bxi} = (n/2)d_t\ln\Tr\boldsymbol{g}^{-1}_\bxi = \grad\cdot\dot{\ex}$.
By identifying the density matrix as similar to the dual metric, $\boldsymbol{g}_\bxi^{-1}$, we can also find this compatibility condition as the equation of motion for the metric determinant, Eq.~\ref{eq:gen_Liouville}, and $\Tr\stability_+$.
Therefore, $|\bxi|^{-\frac{1}{2}}$ obeys the generalized Liouville's equation:
\begin{equation}\label{eq:GLE_xi}
  \frac{\partial}{\partial t}|\bxi|^{-\frac{1}{2}} + \grad\cdot(|\bxi|^{-\frac{1}{2}}\dot{\ex}) = 0,
\end{equation}
which is related to other generalizations~\cite{Tuckerman_1999,TucLiuGioMar2001,Ramshaw_2002,Ezra2004}.
It applies to non-Hamiltonian dynamics but gives the usual Liouville equation if the dynamics are Hamiltonian.
For Hamiltonian dynamics, the metric determinant $|\bxi|^{-1}$ is time-independent, the divergence of the flow vanishes, $d\mathcal{V}(t)=d\mathcal{V}(t_0)$ from Eq.~\ref{eq:invariant_measure}, and the generalized Liouville equation, 
\begin{equation}\label{eq:GLE_xi_Hamiltonian}
\frac{\partial}{\partial t}|\bxi|^{-\frac{1}{2}} + \dot{\ex}\cdot\grad|\bxi|^{-\frac{1}{2}} = 0,
\end{equation}
becomes the usual Liouville equation.

\subsection{Poisson brackets}\label{sec:PB}
\noindent So far, we have analyzed the properties of the determinant of the unnormalized density matrix. However, its trace also appears in the compatibility condition. Analyzing the trace, we find connections to classical dynamics, well-known quantities in dynamical systems, and similarities with quantum mechanics.

Because the dynamics of $\bxi = \sum_{i=1}^{k}\dyad{\delta\bpsi_i}{\delta\bpsi_i}$ are not trace preserving, the rate of change of
$\Tr\bxi$ governs the rate of change of the perturbation,
\begin{equation}
  \frac{1}{2}\frac{d}{dt}\Tr\bxi = \frac{1}{2}\Tr\{\stability_+,\bxi\}
= \Tr(\bxi \stability_+) = \langle \stability_+\rangle_{\bxi}.
\end{equation}
Averages over the unnormalized density matrix are defined akin to quantum mechanical averages; here, the trace evolves at a rate determined by the instantaneous Lyapunov exponents~\cite{PikovskyP16}, measures of local (in)stability, $\langle \stability_+\rangle_{\bxi_i}$. Defining $\langle \stability_+\rangle = \langle \stability_+\rangle_{\bxi}/\Tr\bxi $, the solution to this equation of motion is:
\begin{equation}\label{eq:trace_xi_sol}
  \Tr\bxi(t) = \Tr\bxi(t_0)e^{2\int^t_{t_0}\langle \stability_+(t')\rangle\,dt'}.
\end{equation}
As numerical verification of this result, and all our others, we simulated the dynamics of Hamiltonian and dissipative dynamical systems. As a prototypical dissipative system, we chose the Lorenz model. Figure~\ref{fig:Lorenz_trace_det} shows the time evolution of the trace for a chaotic orbit for 100 random perturbation states drawn from a uniform distribution.
The rapid increase (on the semi-log scale) in $\Tr\bxi$ is indicative of the chaotic nature of the chosen orbit.

For Hamiltonian systems, the equation of motion for the trace can be expressed as a Poisson bracket.
In classical statistical mechanics, a dynamical variable, $f(\boldsymbol{q},\boldsymbol{p})$, can be expressed as $\dot{f}=\{f, \Hamiltonian\}_P$, in terms of the Poisson bracket, $\{.\}_P$.
Combined with our result above, this fact gives a correspondence,
\begin{equation}\label{eq:poisson_connection}
\Tr\dot{\bxi} = 2\langle\stabilityH_+\rangle_\bxi = \{\Tr\bxi, \Hamiltonian\}_P,
\end{equation}
between the Poisson bracket and the tangent-space average of the symmetric part of the stability matrix $\stabilityH_+$.
For example, for an arbitrary perturbation $(u,v)^\top$ in the phase space of the linear harmonic oscillator, $\Tr\,\bxi =u^2+v^2$ and $\Tr\dot{\bxi}=\{\Tr\bxi, \Hamiltonian\}_P = uv(1-\omega^2)$ where $\omega$ is the oscillation frequency.

\begin{figure*}[t]
	\centering
		\resizebox{1.8\columnwidth}{!}
		{
	\begin{minipage}{0.3\textwidth}
		\hspace{-2cm}
		\includegraphics[]{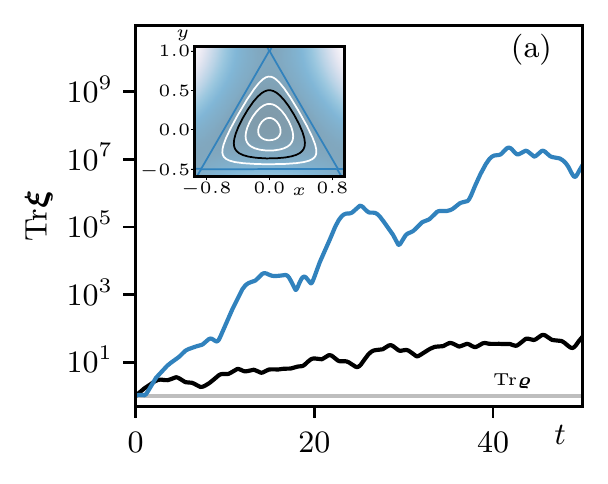}
	\end{minipage}
	\begin{minipage}{.3\textwidth}
		\vspace*{0.7cm}\hspace*{2cm}\small(b)
		
		\hspace{-1.5cm}
		\vspace{0.5cm}
		\resizebox{0.92\columnwidth}{!}
		{
			\begin{tikzpicture}
			\definecolor{S_DBlue}{RGB}{3,78,123}
			\definecolor{blue2}{RGB}{32,99,155}
			\definecolor{S_DGrey}{RGB}{166,166,166}
			\definecolor{S_Gold}{RGB}{255,192,0}
			\begin{axis}
			[
			view={60}{50},
			axis line style={draw=none},
			tick style={draw=none},
			tick label style={color=white}
			]
			\addplot3[
			surf,
			colormap/blackwhite,
			shader=interp,
			domain=-1.2:1.2,
			domain y=-1.13:1.13,
			] 
			{exp(-.15*x^2-.15*y^2)};
			\addplot3[fill=blue2, opacity=0.5]
			coordinates{
				(-1,-1,1)
				(-1,1,1)
				(1,1,1)
				(1,-1,1)
			};
			\end{axis}
			\draw[black,very thick,-latex] (3.6,3.52) -- ++(94:2.5) node[below left] {\large $\grad H$};
			\draw[S_Gold,very thick,-latex] (3.6,3.52) to[bend left] (6.05,2.77);
			\draw[black,very thick,-latex] (3.6,3.52) -- ++(14:2.1);
			\node at (5.1,4.19) {\large\color{black}$\dot{\ex}$};
			\node at (3.52,3.18) {\large$P$};
			\node at (4.8,2.2) {\large$\mathcal{M}$};
			\node at (1.25,4.0) {\large$T\mathcal{M}$};
			\node at (4.8,3.2) {\large\color{S_Gold}$\Gamma(t)$};
			\end{tikzpicture}
		}
	\end{minipage}%
	\begin{minipage}{0.35\textwidth}
		\hspace{-0.8cm}\includegraphics[]{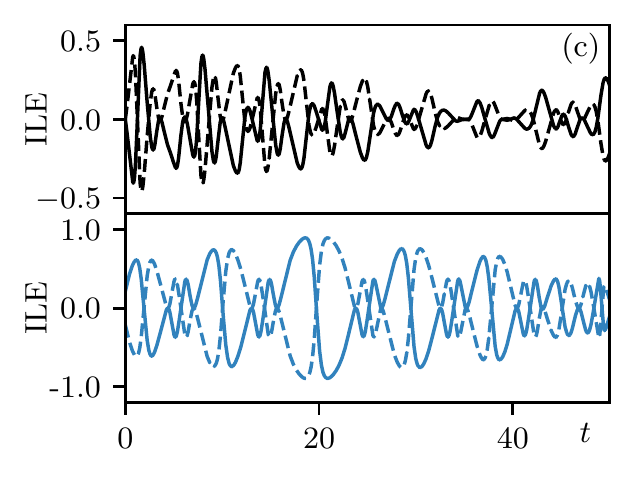}
	\end{minipage}
}
\caption{\label{fig:manifold}(a) The trace of the unnormalized density matrices for a regular (black) and a chaotic orbit (blue) for the H\'enon-Heiles system.
The trace of the normalized density matrices is time invariant (gray).
The inset shows the equipotential surface (and lines) for the potential $U = \frac{1}{2}(x^2+y^2+2x^2y-\frac{2}{3}y^3)$.
Line colors in (a) correspond to those of the equipotential lines.
(b) Schematic illustration of the neighborhood of a point $P$ along a trajectory $\Gamma(t)$ through the phase space $\mathcal{M}$ with the associated tangent space $\mathcal{T}_\mathcal{M}$ and conjugate vectors, $\grad H$ and $\dot{\boldsymbol{x}}$.
(c) Pair of instantaneous Lyapunov exponents (ILE) for the regular with $E=0.0833$ (black) and a chaotic orbit with $E=0.1667$ (blue).}
\end{figure*}
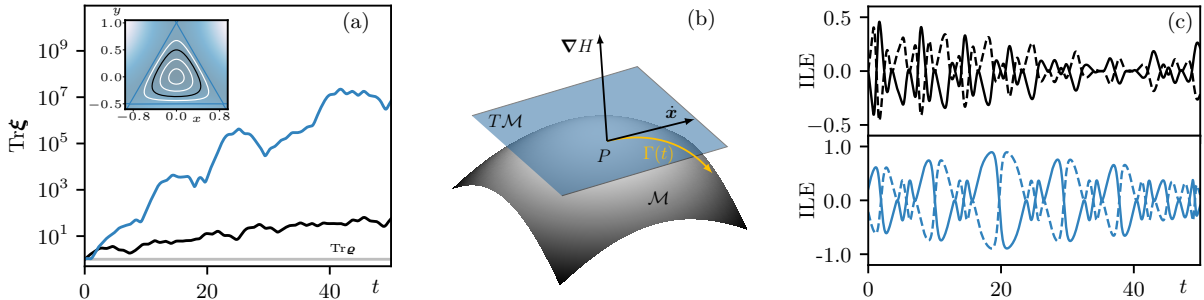

\section{Dynamics of the classical normalized density matrix}\label{sec:dynamics_norm}
\noindent The connections between the density matrix and its properties so far have immediate implications for classical statistical physics. The key results differ in some respects from their analogues in quantum mechanics. For example, the dynamics of $\bxi$ are not norm-preserving, while the norm-preserving dynamics of Hilbert state vectors are a postulate of many formulations of quantum mechanics.

To derive a norm-preserving dynamics for the density matrix of general, classical dynamical systems, we consider a unit perturbation $\ket{\delta \yu} = \ket{\delta \ex}/\|\delta \ex\|$, where $\|.\|$ is the $\ell_2$-norm.
Its equation of motion,
\begin{eqnarray}\label{equ:EOM_unit_per}
  \frac{d}{dt}\ket{\delta \yu} 
&=& (\stability+\stability_-)\,\ket{\delta \yu}- r\,\ket{\delta \yu},
\end{eqnarray}
contains a source/sink term with the instantaneous rate:
$r:=r(t) = \bra{\delta \yu}\stability_+\ket{\delta \yu}= d_t\ln\|\delta \ex(t)\|$.
This rate is the instantaneous Lyapunov exponent for a linearized dynamics, related to the finite-time Lyapunov exponent, $\lambda(t) := \lambda(t,t_0) = |t-t_0|^{-1}\int_{t_0}^tr(t)\,dt$,
and the Lyapunov exponent, $\lambda_i=\lim_{t\to\infty}\lambda_i(t)$. The maximum instantaneous Lyapunov exponent is also referred to as \textit{reactivity} -- the maximum amplification rate over all perturbations, immediately  following the perturbation~\cite{NeuCas1997}.

As before, we represent the state of the dynamical system $\dot{\ex}=\boldsymbol{F}$ as a density matrix. But now, we express it in terms of a unit tangent-space basis $\{\ket{\delta\bphi_i\}}$.
Normalizing each $\ket{\delta\bpsi_i} =c_i\ket{\delta\bphi_i}$, we can define the pure states $\brho_i(t) := \dyad{\delta \bphi_i(t)}{\delta \bphi_i(t)}$ with the expected properties: $\brho_i^2 = \brho_i$, $\Tr\brho_i = 1$, $\Tr\brho_i^2=1$, symmetric, $\brho_i \succeq 1$, i.e., $\brho_i$ is positive semi-definite.
Proving these properties requires the dynamics of $\brho_i$ be norm-preserving.
The perturbation state is the \textit{normalized} density matrix:
\begin{equation}\label{equ:norm_unnorm_den_mat_rel}
  \brho(t) = \frac{\bxi(t)}{\Tr\bxi(t)} = K^{-1}\sum_{i=1}^kc_i^2\dyad{\delta\bphi_i}{\delta\bphi_i},
\end{equation}
which is directly related to the unnormalized $\bxi$ state with $K=\sum_i c_i^2$.
If the state is maximally mixed, with $c_i=1$ $\forall i$ then $\Tr\bxi=k$. 
The tangent-space average $\expval{\brho}{\delta\bphi}$ over the unit basis is analogous to the quantum-mechanical probability of finding the system at $\ket{\delta\bphi}$ given that its state is $\brho$; for example, if  all tangent space directions contribute equally $\expval{\brho}{\delta\bphi}=k^{-1}$.

The normalized state evolves in time according to:
\begin{equation}\label{equ:EOM_rho}
  \frac{d\brho}{dt} = \{\stability_+,\brho\} + [\stability_-,\brho] - 2r\brho,
\end{equation}
another classical analogue of the von Neumann equation in quantum mechanics.
The equation indicates that the average of $\stability_+$ at each instant of time, i.e., the instantaneous Lyapunov exponent is crucial to norm preservation.
It offsets the stretching and contraction of $\ket{\delta \yu}$ due to $\stability$ along a given trajectory. 
Solving this equation of motion, we find the density matrix $\brho$ at time $t_0$ and at a later time $t$ are similar:
\begin{align}\label{eq:rho_sim_trans}
  \brho(t)   = \tilde\jacobian(t,t_0)\brho(t_0)\tilde \jacobian^{-1}(t,t_0),
\end{align}
when $\brho$ is composed of vectors $\{\ket{\delta \yu^i}\}$ evolving in time under $\tilde \jacobian$.
Regardless of the dynamics, the norm-preserving evolution operator $\tilde{\jacobian} := \jacobian\boldsymbol{\Gamma}$ is orthogonal (Appendices~\ref{sec:EOM_LV} and~\ref{sec:proof}) and, so, $|\tilde\jacobian|=1$; the matrix $\boldsymbol{\Gamma}$ has the inverse expansion factors $\|\delta \ex^i(t_0)\|/\|\delta \ex^i(t)\|=e^{-\lambda_i(t) (t-t_0)}$ on the diagonal.
Expressing the unnormalized density matrix $\bxi'=\sum_{i=1}^k\dyad{\delta\ex_i(t)}$ (in terms of the set of vectors that evolve under $\jacobian$), all $\ket{\delta\ex_i(t)}$ collapse on the most rapidly expanding direction.
This collapse motivates well-known methods for computing tangent basis sets~\cite{Benettin1976Kolmogorov,greeneCalculationLyapunovSpectra1987,Wolfe2007,f_ginelli_characterizing_2007} that can also be used to construct density matrices. However, the formalism is not limited to these bases, and, as we will show, can be used to analyze dynamics with other (observable) representations.

\subsection{Generalized Liouville's theorem/equation}\label{sec:GLE_norm}

\noindent Imposing a norm-preserving dynamics not only normalizes the density matrix, it also reinstates the original form of Liouville's equation for non-Hamiltonian dynamics.
To see this, we again take the basis to span the phase space $k\to n$.
As must be the case for a norm-preserving dynamics, the magnitude of perturbations do not evolve with time and $\Tr\brho(t) = \Tr\brho(t_0) = 1$.
The preservation of the trace $\operatorname{Tr}\brho$ follows from Eq.~\ref{equ:EOM_rho}, which shows $\Tr\dot{\brho}=0$ because $\Tr \{\stability_+,\brho\} = 2r$.
However, both the trace and the determinant
\begin{equation}\label{eq:Liouville_rho}
  |\brho(t)| = |\brho(t_0)|,
\end{equation}
are similarity-invariant constants of motion, Figs.~\ref{fig:Lorenz_trace_det} and~\ref{fig:manifold}.
Defining the metric tensor $\boldsymbol{g}_\brho$ such that its determinant $g_\brho = |\brho^{-1}|$, 
the conservation of the determinant is another form of the generalized Liouville's equation, 
\begin{align}
  \frac{\partial}{\partial t}|\brho|^{-\frac{1}{2}} + \dot{\ex}\cdot \grad |\brho|^{-\frac{1}{2}} &= 0.
\end{align}
in terms of $\brho$. The compatibility condition of the metric is 
$\partial_t \sqrt{g_\brho} +  \dot{\ex}\cdot \grad \sqrt{g_\brho} = 0$.

Imposing a norm-preserving dynamics and normalizing the density matrix, however, does not change the form of the generalized Liouville theorem.
With these norm-preserving dynamics, the generalized Liouville theorem in Eq.~\ref{eq:invariant_measure} becomes: $|\boldsymbol{\Gamma}|d\mathcal{V}(t)=|\jacobian||\boldsymbol{\Gamma}| d\mathcal{V}(t_0) =|\tilde\jacobian| d\mathcal{V}(t_0) = d\mathcal{V}(t_0)$ and we can recognize $|\boldsymbol{\Gamma}|^2=|\bxi(t_0)|/|\bxi(t)|$.
Geometrically, the generalized Liouville theorem here is: scaled phase space volumes are conserved under the norm-preserving evolution of a perturbation state with a basis that spans the $n$-dimensional phase space.
This generalization of Liouville's theorem is not limited to Hamiltonian dynamics, however.
In phase space of non-Hamiltonian systems, any part of the initial volume lost (or gained) in course of the dynamics is continually and entirely compensated for by the stretching/contracting of the volume. As a result, the scaled volume $|\boldsymbol{\Gamma}|d\mathcal{V}$ is conserved for the $n$-dimensional phase space of any dynamical system.
For both the harmonic and the damped harmonic oscillator, if the initial volume is set by $|\bxi(t_0)| = 1$ and $\Tr\bxi(t_0)=2$, then $|\brho(t)|=1/4\,\forall t$.
Figure~\ref{fig:manifold}(a) shows $\operatorname{Tr}\bxi$ for the classical H\'enon Heiles system on a regular and chaotic orbit.
\begin{table*}[ht]
	\caption{Equation of motion and instantaneous Lyapunov exponent for the pure state $\brho_i = \dyad{\delta\bphi_i}{\delta\bphi_i}$ in the eigenbases of $\stability_+$, $\stability_-$, and $\stability$, for Hamiltonian and dissipative dynamics.}
	\label{table:summary}
	
	\begin{tabular}{cccc} 
		\hline
		\hline
		& $\stability_+$ & $\stability_-$ & $\stability$ \\ 
		\hline
		\multirow{2}{6.5em}{\textit{Hamiltonian}} & $\dot{\brho_i}= [\stabilityH_-,\brho_i]$ & $\dot{\brho_i}=\{\stabilityH_+,\brho_i\}$& $\dot{\brho_i}=0$\\
		& $r^{\stabilityH_+}_i = \operatorname{Eig}\stabilityH_{+,i}$ & $r_i^{\stabilityH_-} = 0$ & $r_i^{\stabilityH}=\Re(\operatorname{Eig}\stabilityH_i)$ \\
		\hline
		\multirow{2}{6em}{\textit{Dissipative}} & $\dot{\brho_i}= [\stability_-,\brho_i]$ & $\dot{\brho_i}=\{\stability_+,\brho_i\}-2r^{\stability_-}_i\brho_i$& $\dot{\brho_i}=0$ \\
		& $r_i^{\stability_+} = \operatorname{Eig}\stability_{+,i}$ & $r_i^{\stability_-} = \Tr(\stability_+\brho_i)$ & $r^{\stability}_i=\Re(\operatorname{Eig}\stability_i)$\\
		\hline
		\hline
	\end{tabular}
\end{table*}

\subsection{Basis representation} \label{sec:basis}
\noindent In quantum statistical mechanics, the choice of basis states provides an explicit matrix representation of the quantum state.
Here, there are also different basis sets that can be used for perturbation states: $\brho$ and $\bxi$ (Appendix~\ref{sec:basis_rep}).
In dynamical systems theory, it is common to analyze Lyapunov vectors, such as Gram-Schmidt vectors~\cite{Benettin1976Kolmogorov,shimada_numerical_1979}.
More recently, however, there has been an interest in covariant Lyapunov vectors~\cite{f_ginelli_characterizing_2007}.
Based on early work~\cite{DellagoPH96}, Lyapunov vectors with small, but finite, exponents are hydrodynamic modes that characterize macroscopic transport~\cite{McNamaraM01,YangR05}.

The choice of basis has consequences for the form of the equation of motion for the density matrix (Appendix~\ref{sec:basis_rep} and Table~\ref{table:summary}).
A natural basis that spans the phase space is the eigenvectors of $\stability_+$ because the instantaneous Lyapunov exponents are the associated eigenvalues.
In this basis, the equation of motion for each $\brho_i=\dyad{\delta\bphi_i}{\delta\bphi_i}$ reduces to the commutator $\dot\brho_i = [\stability_-,\brho_i]$.
However, choosing the eigenvectors of $\stability_-$ as the basis commutator vanishes and $\dot\brho_i = \{\stability_+,\brho_i\} - 2r_i^{\stability_-}\brho_i$.
The eigenbasis of $\stability$ appears to be unique in that $\brho$ is time independent.
Hamiltonian dynamics with $\Hamiltonian(q^i, p_i)\in\mathbb{R}$ lead to interesting consequences for these basis-dependent results (Appendix~\ref{sec:DMD_Hamiltonian}).
For example, in the $\stabilityH_-$ basis: $\dot\brho_i = \{\stabilityH_+,\brho_i\}$.
But, in the $\stabilityH$ basis, $\Tr\bxi_i$ is a constant of motion, $\{\Tr\bxi_i,\Hamiltonian\}_P=0$.

Hamiltonian systems also have special tangent space directions associated with conserved quantities. On a constant energy manifold, for example, there are two conjugate tangent directions: the phase velocity $\dot \ex$ and the gradient of the Hamiltonian $\grad\Hamiltonian$, Fig.~\ref{fig:manifold}(b).
They are related $\dot \ex=\bm{\Omega} \grad\Hamiltonian$ through the Poisson matrix $\bm{\Omega}$, orthogonal to each other, $\grad\Hamiltonian\cdot \bm{\Omega} \grad \Hamiltonian = 0$, and have equal magnitude $\|\dot \ex\|=\|\grad \Hamiltonian\|$ through Hamilton's equations.
The vector $\grad\Hamiltonian$ is also orthogonal to the constant energy manifold and used to define the invariant measure~\cite{dorfman1999introduction}.
In general, the vector sets defining the density matrices here need not span the whole phase space.
So, forming density matrices from each conjugate vector, we find the pure states, $\brho_{\dot \ex}$ and $\brho_{\bm{\Omega}\grad \Hamiltonian}$, both have unit trace and are related by $\brho_{\dot \ex} = \bm{\Omega}\brho_{\grad \Hamiltonian}\bm{\Omega}^\top$ (Appendix~\ref{sec:DMD_Hamiltonian}). The conjugate pure states in the tangent space for a 2D-Hamiltonian system are:
\begin{align}
\|\dot \ex\|^2\brho_{\dot \ex} &=
\begin{pmatrix}
\dot q^2 & \dot p\dot q \\[2pt]
\dot p\dot q & \dot p^2\\ 
\end{pmatrix}, \nonumber\\
\|\grad\Hamiltonian\|^2\brho_{\grad\Hamiltonian} 
&=
\begin{pmatrix}
\dot p^2 & -\dot p\dot q\\
-\dot p\dot q & \dot q^2
\end{pmatrix}.\label{eq:tangent_pure_states}
\end{align}
These states are formed from the outer product of the unit tangent vectors, $\ket{\delta\bphi_{\dot{\ex}}} = \|\dot{\ex}\|^{-1}(q,p)^\top$ and $\ket{\delta\bphi_{\grad{\Hamiltonian}}} = \|\grad\Hamiltonian\|^{-1}(-\dot{p},\dot{q})^\top$,
where $\|\dot \ex\|^2 = \|\grad\Hamiltonian\|^2 = \dot{p}^2 + \dot{q}^2$.

These particular density matrices have an interesting (lower dimensional) parallel with Liouville's theorem. Liouville's theorem can be thought of as an equivalence of the divergence of the phase flow, the trace of the Jacobian, and the intrinsic rate:
$d_t\ln \delta\mathcal{V}=\grad\cdot\dot{\ex} = n\Tr[\brho\stabilityH(\boldsymbol{x})]$,
for $\brho = n^{-1}\sum_{i=1}^{n}\dyad{\delta\bphi_i}{\delta\bphi_i}$. There is a similar equivalence for the instantaneous Lyapunov exponents of $\dot\ex$ and $\grad \Hamiltonian$.
For a $2D$-Hamiltonian system, these take the form:
\begin{align}\nonumber
\frac{d\ln\|\dot \ex\|}{d t} \nonumber &= \frac{\dot q\cdot\dot p}{\|\dot \ex\|^2}\grad\cdot 
\begin{pmatrix} \dot p \\ \dot q \end{pmatrix} = \Tr (\brho_{\dot x}\stabilityH_+), \\
\frac{d \ln\|\grad \Hamiltonian\|}{d t}  &= \frac{-\dot q\cdot\dot p}{\|\grad \Hamiltonian\|^2}\grad\cdot 
\begin{pmatrix} \dot p \\ \dot q \end{pmatrix} = \Tr(\brho_{\grad H}\stabilityH_+),
\end{align}
denoting the phase point $\ex=(q,p)$.
Unlike, the intrinsic rate of the volume element $\delta\mathcal{V}=\delta q\delta p$ that appears in Liouville's theorem, these instantaneous Lyapunov exponents are not zero -- they are related to the divergence in a common direction $(\dot p, \dot q)^\top$ that is a reflection about $\dot p= \dot q$.
Instead, they are conjugate, so they sum to zero and span an area that is conserved.

\section{Case studies}\label{sec:case_studies}

To illustrate this density matrix formalism and show that the main ingredients of the theory are computable, we apply it analytically to the linear and damped harmonic oscillator and numerically to the H\'enon-Heiles and Lorenz models (Appendix~\ref{sec:case_examples}).
Below, we explicitly derive the instantaneous Lyapunov exponents of the damped and simple harmonic oscillator in special tangent space directions.
Analytical expressions for these exponents in the eigenvector bases of the stability matrix $\stability$ are summarized in Table~\ref{table:HOsummary}.

\begin{table*}[ht]
	\caption{Instantaneous Lyapunov exponents for the linear and damped harmonic oscillator in the eigenvector basis of the stability matrix $\stability$ and its (anti)symmetric parts. The harmonic oscillator with frequency $\omega$ has the Hamiltonian: $H = p^2 + \omega^2q$; in this case, $\{\Tr \brho,H\} = 0$. The equations of motion for the damped harmonic oscillator with the dissipation parameter $\gamma$ are $\dot{q} = p$ and $\dot{p} = -\omega^2q - \gamma p$.}
	\label{table:HOsummary}
	
	\begin{tabular}{cccc} 
		\hline
		\hline
		Harmonic oscillator & $\stability_+$ & $\stability_-$ & $\stability$\\ 
		\hline
		\vspace{2pt}
		\textit{Linear} & $r_i^{\stabilityH_+} = \pm (1-\omega^2)/2$&
		\vspace{2pt}
		$r_i^{\stabilityH_-} = 0$&
		$r_i^{\stabilityH} = 0$\\[2pt]
		\textit{Damped} & $r_i^{\stability_+} = \frac{1}{2}(-\gamma \pm \sqrt{\gamma^2 + (1-\omega^2)^2})$ &
		$r_i^{\stability_-} = -\frac{1}{2}\gamma$ &
		$r_i^{\stability} = \frac{1}{2}(-\gamma \pm \Re (\sqrt{\gamma^2 - 4\omega^2}))$\\[2pt]
		\hline
		\hline
	\end{tabular}
\end{table*}
\medskip

\noindent\textit{a. Linear harmonic oscillator.--} The simple one-dimensional harmonic oscillator for unit mass and frequency $\omega$ has the Hamiltonian $\Hamiltonian = \frac{1}{2}(p^2 + \omega^2 q^2)$. From the usual equations of motion, $\dot{q} = p$, and $\dot{p} = -\omega^2\,q$, the stability matrix $\stabilityH$ is time independent,
\begin{align}
\stabilityH &= \left(
\begin{array}{cc}
0 & 1\\
-\omega^2 & 0\\
\end{array}\right),
\end{align}
with symmetric and anti-symmetric parts:
\begin{align}
\stabilityH_+&= \frac{1}{2}\left(
\begin{array}{cc}
0 & 1-\omega^2\\
1-\omega^2 & 0\\
\end{array}\right), \nonumber\\
\stabilityH_-&= \frac{1}{2}\left(
\begin{array}{cc}
0 & 1+\omega^2 \\
-1-\omega^2 & 0\\
\end{array}\right).
\end{align}
In the basis of $\stabilityH$ and $\stabilityH_-$, the instantaneous Lyapunov exponents vanish, Appendix~\ref{sec:DMD_Hamiltonian}.
Only the instantaneous Lyapunov exponents in the $\stabilityH_+$ eigenvector basis is nonzero, Table~\ref{table:HOsummary}. Because the dynamics are Hamiltonian, there is a conjugate pair of exponents with a magnitude that depends on the oscillator frequency $\omega$.
Appendix Figs.~\ref{fig:SI_HO}(a) and~\ref{fig:SI_HO}(b) show the time evolution of the instantaneous Lyapunov exponents and $\Tr\bxi$ for an arbitrary pure state (formed from a random unit perturbation vector) on a sampled trajectory.

The instantaneous Lyapunov exponents in the conjugate tangent space directions $\dot\ex$ and $\grad H$ for the linear harmonic oscillator are also a conjugate pair. From Eq.~\ref{eq:tangent_pure_states}, the pure states are:
\begin{align}
\brho_1 &=  \|\dot \ex\|^{-2}
\begin{pmatrix}
p^2 &	-\omega^2pq\\ 
-\omega^2pq &	\omega^4q^2
\end{pmatrix}\nonumber \\
\brho_2 &=  \|\grad\Hamiltonian\|^{-2}\begin{pmatrix}
\omega^4q^2 &	\omega^2pq\\ 
\omega^2pq & p^2\\
\end{pmatrix},
\end{align}
where $\|\dot \ex\|^{2} = \|\grad\Hamiltonian\|^2 = p^2 + \omega^4q^2$. The instantaneous Lyapunov exponents for these states,
\begin{align}\label{eq:ILE_HO}
r_{1,2} & = \Tr(\stabilityH_{+}\brho_{1,2})= \pm (1-\omega^2)\omega^2\frac{pq}{p^2+\omega^4q^2},
\end{align}
are functions of time, as shown in Appendix Fig.~\ref{fig:SI_HO}(c) using a frequency of $\omega = 0.5$.
In the eigenbasis of $\stabilityH_+$, the instantaneous Lyapunov exponents are (from Appendix Eq.~\ref{eq:ILE_symA}): $r^{\stabilityH_+}_{1,2} = \pm \frac{1}{2}(1-\omega^2)$.
Therefore, the instantaneous Lyapunov exponents in these special directions are the eigenvalues of $\stabilityH_+$ scaled by a function of the state. 

These results can be generalized to higher dimensions, as we show for the H\'enon-Heiles system in Appendix~\ref{sec:case_examples}. Figure~\ref{fig:manifold}(c) shows the conjugate pair of instantaneous Lyapunov exponents for the H\'enon Heiles system.

\smallskip

\noindent\textit{b. Damped harmonic oscillator.--} For the damped harmonic oscillator and a velocity dependent dissipation parameterized by $\gamma$, the equations of motion are:
\begin{align}\label{eq:DHO}
\dot{q} &= p  \nonumber\\
\dot{p} &= -\omega^2 q - \gamma p.
\end{align} 
Conservative dynamics are recovered for $\gamma = 0$.
The stability matrix of this system,
\begin{align}
\stability =\left(
\begin{array}{cc}
0 & 1\\
-\omega^2 & -\gamma\\
\end{array}\right),
\end{align}
has symmetric and anti-symmetric parts:
\begin{align}\nonumber
\stability_+ &=\frac{1}{2}\left(
\begin{array}{cc}
0 & 1-\omega^2\\
1-\omega^2 & -2\gamma\\
\end{array}\right), \\
\stability_- &=\frac{1}{2}\left(
\begin{array}{cc}
0 & 1+\omega^2\\
-1-\omega^2 & 0\\
\end{array}\right). 
\end{align}
Appendix Figs.~\ref{fig:SI_DHO}(a) and~\ref{fig:SI_DHO}(b) show the time evolution of $\Tr\bxi$ and instantaneous Lyapunov exponent for a pure state given by a random unit perturbation vector. 

The instantaneous Lyapunov exponents in the conjugate tangent space directions  for the damped harmonic oscillator are found, as before, from the pure states in Eq.~\ref{eq:tangent_pure_states}:
\begin{align}
\brho_1 &=  \|\dot \ex\|^{-2}
\begin{pmatrix}
p^2 &	-(\omega^2q +\gamma p)p\\ 
-(\omega^2q +\gamma p)p &	(\omega^2q + \gamma p)^2
\end{pmatrix}\nonumber \\
\brho_2 &=  \|\grad\Hamiltonian\|^{-2}\begin{pmatrix}
(\omega^2q + \gamma p)^2 &	(\omega^2q +\gamma p)p\\ 
(\omega^2q +\gamma p)p & p^2\\
\end{pmatrix},
\end{align}
where $\|\dot \ex\|^{2} = \|\grad\Hamiltonian\|^2 = p^2 + (\omega^2q + \gamma p)^2$. The instantaneous Lyapunov exponent for these states,
\begin{align}
r_{1} & = \Tr(\stability_{+}\brho_{1})\nonumber\\
&= \frac{-(\omega^2q + \gamma p)}{p^2+(\omega^2q + \gamma p)^2}\Big[(1-\omega^2)p+ \gamma(\omega^2q+\gamma p)\Big], \nonumber\\[5pt]
r_{2} & = \Tr(\stability_{+}\brho_{2})\nonumber\\
& = \frac{\omega^2p}{p^2+(\omega^2q + \gamma p)^2}\Big[(1-\omega^2)q - \gamma p\Big].
\end{align}
are shown in Appendix Fig.~\ref{fig:SI_DHO}(c) using a frequency of $\omega = 0.5$ and $\gamma=0.05$.
Note that Eq.~\ref{eq:ILE_HO} is recovered when $\gamma = 0$.

We can also compute these instantaneous Lyapunov exponents in the other eigenbases. For instance, in the $\stability_+$-basis, its eigenvalues are:
\begin{align}
r^{\stability_+}_{1,2} &= \frac{1}{2}(-\gamma \pm \sqrt{\gamma^2 + (1-\omega^2)^2}).
\end{align}
In the $\stability$ eigenbasis, these are $r_i^{\stability} = \Re(\lambda^\stability_{i})$ (using Appendix  Eq.~\ref{eq:ILE_A}),
\begin{align}
r_i^{\stability} = \Re \lambda^\stability_i = \Re \frac{1}{2}(-\gamma \pm \sqrt{\gamma^2 - 4\omega^2}),
\end{align} 
where $\lambda^\stability_i$ indicates the eigenvalue of $\stability$. 
The real part of $\lambda^\stability_i$ gives the instantaneous Lyapunov exponent in the $\stability$ basis. These exponents are unequal when $\gamma^2>4\omega^2$. 
In the $\stability_-$ basis, we can compute the instantaneous Lyapunov exponents as an expectation value over the normalized density matrix, $r_i^{\stability_-} = \Tr (\stability_-\brho_i)$, where each $\brho_i$ is the eigenvectors of $\stability_-$ (see Appendix Eq.~\ref{eq:basis_states_A_anti}).
Doing so, we find the instantaneous Lyapunov exponents are directly proportional to the damping coefficient $r_i^{\stability_-} = -\gamma/2$. 

These results for a dissipative system can be also generalized to higher dimensions, as we show for the Lorenz model in Appendix~\ref{sec:case_examples}.
Appendix Fig.~\ref{fig:SI_Lorenz} shows the instantaneous Lyapunov exponents for the Lorenz model.

\section{Conclusions}\label{sec:conclusions}

\noindent Liouville's equation and theorem are the foundation of statistical mechanics established by Gibbs, Maxwell, and Boltzmann.
Boltzmann, for example, approximated Liouville's equation to derive his $H$-theorem for irreversible processes, making an assumption of ``molecular chaos''.
Here, we have established a density matrix formulation of dynamical systems that explicitly and quantitatively accounts for measures of local instability and chaos -- Lyapunov exponents.
Through this connection, we could derive generalizations Liouville's theorem/equation for any differentiable dynamical system.
And, when the dynamics are Hamiltonian, these generalizations reduce to the traditional forms of the Liouville theorem and Liouville's equation.
We have shown they derive from the properties of classical density matrices, which themselves evolve under an equation of motion akin to the von Neumann equation at the foundation of quantum statistical mechanics.
From these results, the generalized Liouville equation becomes numerically computable and, thus, a new basis for analyzing classical speed limits on observables~\cite{das2021speed}, the spread of perturbations, and the transport of statistical density in the living, synthetic, and engineered dynamical systems across physics.

\begin{acknowledgments}
This material is based upon work supported by the National Science Foundation under Grant No. 2124510 and 1856250.
This publication was also made possible, in part, through the support of a grant from the John Templeton Foundation.
\end{acknowledgments}
\appendix

\section{Equation of motion for Lyapunov vectors}\label{sec:EOM_LV}

Consider the time evolution equation of a generic, infinitesimal perturbation $\ket{\delta \ex}\in T\mathcal{M}$,
\begin{equation}
\frac{d}{dt}\ket{\delta\ex(t)} = \stability(\ex)\ket{\delta \ex(t)},
\label{equ:EOM_per_ket}
\end{equation}
governed by the stability matrix of the system, $\stability$.
Using the time-ordering operator $\mathcal{T}_+$, perturbations propagate as:
\begin{align}\label{equ:per_ket}
\ket{\delta \ex(t)} &= \jacobian(t,t_0) \ket{\delta \ex(t_0)} \nonumber \\
&= \mathcal{T}_+e^{\int_{t_0}^t\stability(t')\,dt'}\ket{\delta \ex(t_0)}.
\end{align}
The evolution operator or Jacobian matrix $[\jacobian(t,t_0)]^i_{\,j} = \partial x^i(t)/\partial x^j(t_0)$ has the equation of motion:
\begin{equation}
\frac{d\jacobian}{dt} = \stability\jacobian
\quad\quad\text{or}\quad\quad
\stability = \frac{d\jacobian}{dt}\jacobian^{-1}.
\end{equation}
The determinant obeys Jacobi's formula:
\begin{equation}
\frac{d}{dt} |\jacobian|=|\jacobian| \Tr\left(\frac{d\jacobian}{dt}\jacobian^{-1}\right)=|\jacobian| \Tr\stability.
\end{equation}
Each $\ket{\delta \ex}\in T\mathcal{M}$ has a corresponding $\bra{\delta \ex}\in T\mathcal{M}$ $\bra{\delta \ex}\in T\mathcal{M}$.
To find the dynamics of the dual vector $\bra{\delta \dot{\ex}(t)}$, we partition the stability matrix $\stability = \stability_+ + \stability_-$ into its symmetric and anti-symmetric parts, $\stability_\pm = \frac{1}{2}(\stability \pm \stability^\top)$.
Dual vectors then evolve according to:
\begin{equation}\label{equ:EOM_bra}
\frac{d}{dt}\bra{\delta \ex(t)} = \bra{\delta \ex(t)}(\stability_{+}-\stability_-).
\end{equation}
Together, the equations for the motion of tangent vectors and their dual define the non-unitary dynamics of $\ket{\delta \ex}$ in the tangent space, $\braket{\delta \ex(t)}{\delta \ex(t)} \neq \braket{\delta \ex(t_0)}{\delta \ex(t_0)}$.

The time evolution of a unit Lyapunov vector $\ket{\delta \yu}$ in the phase space of a dynamics system,
\begin{align}\label{eq:ket_rho}
\frac{d}{dt}\ket{\delta \yu} =\stability_+\ket{\delta \yu} + \stability_-\ket{\delta \yu} - r \ket{\delta \yu},
\end{align}
has an additional source/sink term with the instantaneous Lyapunov exponent $r = \langle \stability_+\rangle = \bra{\delta \yu}\stability_+\ket{\delta \yu}$.
The solution is:
\begin{equation}
\ket{\delta \yu} = \left(\jacobian{\bf\Gamma}\right) \ket{\delta \yu(t_0)} =: \tilde{\jacobian}(t,t_0)\ket{\delta \yu(t_0)}.
\end{equation}
For any dynamics through the state space, the norm-preserving evolution operator $\tilde{\jacobian} := \jacobian\boldsymbol{\Gamma}$ is orthogonal (Appendix~\ref{sec:proof}) and, so, $|\tilde\jacobian|=1$. The matrix $\boldsymbol{\Gamma}$ has the inverse expansion factors $\|\delta \ex_i(t_0)\|/\|\delta \ex_i(t)\|=e^{-\lambda_i(t) (t-t_0)}$ on the diagonal.
The equation of motion for $\bra{\delta \yu}$,
\begin{align}
\frac{d}{dt}\bra{\delta \yu} = \bra{\delta \yu}\stability_+ - \bra{\delta \yu}\stability_- - \langle \stability_+\rangle \bra{\delta \yu},
\end{align}
has the solution:
\begin{equation}
\bra{\delta \yu} = \bra{\delta \yu(t_0)}\left(\jacobian{\bf\Gamma}\right)^\top =: \bra{\delta \yu(t_0)}\tilde{\jacobian}^\top(t,t_0).
\end{equation}

The equation of motion for density matrices follow from these results for tangent vectors.
In the main text, we consider basis sets that span the $n$-dimensional phase space, defining the unnormalized,
\begin{equation}
  \boldsymbol{\xi} = \sum_{i=1}^{k}\dyad{\delta \bpsi_i} = \sum_{i=1}^{k}c_i^2\dyad{\delta \boldsymbol{\phi}_{i}},
\end{equation}
and normalized
\begin{equation}
\brho(t) = K^{-1}\sum_{i=1}^k c_i^2\brho_i = K^{-1}\sum_{i=1}^kc_i^2\dyad{\delta\bphi_i}{\delta\bphi_i},
\end{equation}
density matrices. Here, $K=\Tr \bxi =\sum_{i=1}^kc_i^2$.
They evolve according to Eq.~\ref{eq:EOM_xi} and Eq.~\ref{equ:EOM_unit_per} in the main text. Each pure state $\brho_i$ evolves as:
\begin{align}\nonumber
\frac{d\brho_i}{dt}&=\frac{d}{dt}(\dyad{\delta \bphi_i})\\
&= \ket{\delta \bphi_i}\left(\frac{d}{dt}\bra{\delta \bphi_i}\right) +\left(\frac{d}{dt}\ket{\delta \bphi_i}\right) \bra{\delta \bphi_i} \nonumber\\
&= \brho_i\stability_+ - \brho_i\stability_-+ \stability_+\brho_i + \stability_-\brho_i - 2r\brho_i \nonumber\\
& = \{\stability_+,\brho_i\} + [\stability_-,\brho_i] -2r_i\brho_i.
\end{align}

\section{Generalized Liouville theorem and equation}\label{sec:GLE}

For the unnormalized density matrix $\bxi$,
\begin{equation}
\ln |\bxi| = \Tr\ln \bxi,
\end{equation}
follows from the identity $\ln |\boldsymbol{C}| = \Tr\ln \boldsymbol{C}$ between the trace and the determinant $|.|$. 
Assuming $\bxi$ is invertible, taking the time derivative,
\begin{align}
\frac{d}{dt}\ln |\bxi| &= \Tr \frac{d}{dt}\ln \bxi= \Tr \bxi^{-1}\frac{d\bxi}{dt},
\end{align}
and using Eq.~\ref{eq:EOM_xi}, we find the generalized Liouville's equation in the main text:
\begin{align}
\frac{d}{dt}\ln|\bxi| &= \Tr \bxi^{-1}(\{\stability_+,\bxi\}+[\stability_-,\bxi]) = 2\Tr\stability_+.
\end{align}
To see that this result is a generalization of Liouville's equation requires recognizing the phase space volume element is $d\mathcal{V} = dx^{1} \wedge \cdots \wedge d x^{n}$.
The determinant of the Jacobian $\jacobian$ governs its coordinate transformation under the action of the dynamical equations: $d\mathcal{V}(t)=| \jacobian(t,t_0)|d\mathcal{V}(t_0)$.
Equation~\ref{eq:invariant_measure} follows from the determinant of Eq.~\ref{eq:op_M}: $|\boldsymbol{\xi}(t)|=|\boldsymbol{M}\left(t, t_{0}\right)|^2 |\boldsymbol{\xi}\left(t_{0}\right)|$.

Treating the phase space as a general Riemannian manifold endowed with a (contravariant) metric tensor $\boldsymbol{g}_\bxi^{-1}$, we identify this metric tensor as \emph{similar} to the unnormalized density matrix $\bxi$.
To see this relationship, consider two arbitrary ordered bases ${\ket{\delta\bpsi_i(t_0)}}$ and ${\ket{\delta\bpsi_i(t)}}$ stacked in matrix columns,
\begin{align}
  \bPsi' := \bPsi(t_0) &= \Big[\ket{\delta\bpsi_1(t_0)}, \cdots, \ket{\delta\bpsi_n(t_0)}\Big] \\\bPsi(t) &= \Big[\ket{\delta\bpsi_1(t)}, \cdots, \ket{\delta\bpsi_n(t)}\Big],
\end{align}
and related by $\bPsi'= \jacobian\bPsi$.
In these bases, we can represent the transformation of the unnormalized density matrix:
\begin{align}
  \bxi(t) &= \bPsi'\bPsi'^\top  = \boldsymbol{M}\bPsi \bPsi^\top\boldsymbol{M}^\top = \boldsymbol{M}\bxi(t_0) \boldsymbol{M}^\top.
\end{align}

However, the linear transformation $\bPsi\to \bPsi'$ is also obtained by a change-of-basis matrix $\boldsymbol{P}$ as $\bPsi'=\bPsi\boldsymbol{P}$.
The Jacobian $\jacobian$ and $\boldsymbol{P}$ are related by a similarity transformation
\begin{align}
  \boldsymbol{P} = \bPsi^{-1}\jacobian\bPsi.
\end{align}
One can then view $\boldsymbol{P}$ and $\jacobian$ as \emph{propagators} expressed in different bases that represent the linear transformation of $\bPsi\to \bPsi'$ forward in time.
The Jacobian matrix $\jacobian$ comes with a natural co-ordinate basis, $\{\partial/\partial x^i\}$.
By constructing another, more convenient basis, through the density matrix the dynamics are governed by $\boldsymbol{P}$.
That $\boldsymbol{P}$ is similar to $\jacobian$, implies $\boldsymbol{P}\to\jacobian$ when the density matrix is expressed in the coordinate basis.
The contravariant metric tensor $\boldsymbol{g}^{-1}_\bxi$ transform as:
\begin{align}
  \boldsymbol{g}^{-1}_\bxi(t) = \bPsi'^\top \bPsi' = \boldsymbol{P}^\top \bPsi^\top \bPsi\boldsymbol{P} = \boldsymbol{P}^\top \boldsymbol{g}_\bxi^{-1}(t_0)\boldsymbol{P},
\end{align}
and the covariant metric tensor $\boldsymbol{g}_\bxi(t)$ transforms as:
\begin{align}
  \boldsymbol{g}_\bxi(t) = \boldsymbol{P}^{-1} \bm{g}_\bxi(t_0)\boldsymbol{P}^{-\top}.
\end{align}
It follows from these relationships that $\bxi$ and $\bm{g}_\bxi^{-1}$ are similar:
\begin{align}\label{eq:xi_g_sym}
  \bPsi'^{-1}\,\bxi(t)\,\bPsi' =  \bPsi'^{-1}\,\bPsi'\bPsi'^\top\,\bPsi' = \bm g^{-1}_\bxi(t). 
\end{align}
If $\bPsi'=\jacobian\bPsi = \bPsi\boldsymbol{P}$ with $\boldsymbol{P}=\bPsi^{-1} \jacobian \bPsi$ then
\begin{align}
    \boldsymbol{g}_\bxi(t)^{-1} &= \bPsi'^{-1}\,\bxi(t)\,\bPsi' = \boldsymbol{P}^{-1} \boldsymbol{g}^{-1}_\bxi(t_0)\boldsymbol{P}.
\end{align}
The linear independence of the basis vectors in $\bPsi$ guarantees it is invertible.

For a general Riemannian manifold, the volume $n$-form determines the \textit{invariant} volume element $d\tilde{\mathcal{V}}$ in an arbitrary coordinate system: $d\tilde{\mathcal{V}}=\sqrt{g_\bxi}d\mathcal{V}$, where $g_\bxi$ is the determinant of the covariant metric tensor $\boldsymbol{g}_\bxi$.
From Eq.~\ref{eq:xi_g_sym}, $\boldsymbol{g}_\bxi$ is similar to $\bxi^{-1}$.

Furthermore, Eq.~\ref{eq:gen_Liouville} provides the compatibility condition of the metric tensor with the flow:
\begin{align}
\frac{d}{dt}\ln|\bxi| &= 2\,\grad\cdot\dot{\ex} \nonumber\\
\frac{d}{dt}|\bxi|&= 2|\bxi|\grad\cdot\dot{\ex} \nonumber\\
\frac{1}{2}|\bxi|^{-\frac{3}{2}}\frac{d}{dt}|\bxi| &= |\bxi|^{-\frac{1}{2}}\grad\cdot\dot{\ex} \nonumber\\
\frac{d}{dt}|\bxi|^{-\frac{1}{2}}&=-|\bxi|^{-\frac{1}{2}}\grad\cdot\dot{\ex}.
\end{align}
We can express this relation in terms of the metric determinant~\cite{Tuckerman_1999}:
\begin{align} \nonumber
\frac{d}{dt}\sqrt{g_\bxi} &= -\sqrt{g_\bxi}\grad\cdot\dot{\ex}\\ \nonumber
\frac{\partial\sqrt{g_\bxi}}{\partial t}+\dot{\ex}\cdot \grad\sqrt{g_\bxi} &=-\sqrt{g_\bxi} \grad\cdot\dot{\ex}\\ \nonumber
\frac{\partial}{\partial t}\sqrt{g_\bxi}+\grad\cdot(\sqrt{g_\bxi}\dot{\ex}) &=0.
\end{align}
Replacing $g_\bxi$ by $|\bxi|^{-1}$, we obtain the generalized Liouville equation:
\begin{equation}
\frac{\partial}{\partial t}(|\bxi|^{-\frac{1}{2}}) + \grad\cdot(|\bxi|^{-\frac{1}{2}}\dot{\ex}) = 0.
\end{equation}
For Hamiltonian dynamics, the metric is time-independent because of the vanishing flow divergence,
\begin{equation}
\frac{\partial}{\partial t}(|\bxi|^{-\frac{1}{2}}) + \dot{\ex}\cdot\grad|\bxi|^{-\frac{1}{2}} = 0.
\end{equation}

By introducing a norm-preserving dynamics and a normalized density matrix, the form of the generalized Liouville equation is identical to the traditional Liouville equation.
Defining the metric determinant $g_\brho=|\boldsymbol{\brho}|^{-1}$, the conservation of the normalized density matrix $\brho$,
\begin{align}\nonumber
\frac{d}{dt}\ln|\brho| &= 0,\nonumber
\end{align}
is equivalent to the generalized Liouville equation:
\begin{align}
\frac{d}{dt}(|\brho|^{-\frac{1}{2}}) = \frac{d}{dt}\sqrt{g_\brho} &= 0 \nonumber\\
\frac{\partial}{\partial t}(|\brho|^{-\frac{1}{2}}) + \dot{\ex}\cdot\grad|\brho|^{-\frac{1}{2}} &= 0 \\ \frac{\partial}{\partial t}(\sqrt{g_\brho}) + \dot{\ex}\cdot\grad\sqrt{g_\brho} &=0.
\end{align}

The Liouville equations for $\bxi$ and $\brho$ are related.
Taking the determinant of Eq.~\ref{equ:norm_unnorm_den_mat_rel}, we find
\begin{align}\nonumber
|\brho| &= \frac{|\bxi|}{(\Tr\bxi)^n} \\\nonumber
g_\brho &= g_\bxi\,(\Tr\bxi)^n\\\nonumber
\sqrt{g_\brho} &= \sqrt{g_\bxi}\,(\Tr\bxi)^{\frac{n}{2}}\\
\ln\sqrt{g_\brho} &= \ln\sqrt{g_\bxi}+\frac{n}{2}\ln(\Tr\bxi).
\end{align}
Recalling that $g_\brho$ is time-independent, the ratio $|\bxi|/\Tr\bxi^n$ is a constant of motion for any dynamical system.

It is also possible to express the compatibility condition using the trace of $\bm{g}^{-1}_\bxi$,
\begin{align}
-\frac{d}{dt}\ln\sqrt{g_\bxi} = \frac{n}{2}\frac{d}{dt}\ln(\Tr\bar{\boldsymbol{g}}_\bxi)=\grad\cdot\dot{\ex}.
\end{align}
For the normalized density matrix $\brho$ of the form $\brho(t) = n^{-1}\sum_{i=1}^{n}\brho_i(t)$, we have averages similar to those in quantum mechanics. For example, the average
\begin{align}
\Tr(\stability_+\brho) &= n^{-1}\sum_{i=1}^{n}\Tr(\stability_+\brho_i) \nonumber\\
&= n^{-1}\sum_{i=1}^{n}r_i = n^{-1}\Tr\stability_+,
\end{align}
where $r_i$ is the instantaneous Lyapunov exponent for the $i^\textrm{th}$ basis state.
This average is related to the divergence of the flow: $\grad\cdot\dot{\ex} = n\Tr(\stability_+\brho)$.

\section{Proof that $\tilde{\jacobian}$ is orthogonal} \label{sec:proof}

The norm-preserving evolution operator is: $|\tilde{\jacobian}| = |\boldsymbol{\Gamma}||\jacobian|$. Applying the identity $|e^{\stability}|=e^{\Tr\stability}$ to the determinant of the Jacobian gives:
\begin{equation}
  |\tilde{\jacobian}| = |\boldsymbol{\Gamma}||\jacobian| = |\boldsymbol{\Gamma}|e^{\Tr\int_{t_0}^t \stability\,dt'} = |\boldsymbol{\Gamma}|e^{\int_{t_0}^t \Tr\stability\,dt'}.
\end{equation}
Using the fact that $\boldsymbol{\Gamma}$ is diagonal and that trace of a matrix is the sum of eigenvalues:
\begin{equation}
  |\tilde{\jacobian}| = e^{-\int_{t_0}^t\sum\limits_i^n\lambda_idt'}e^{\int_{t_0}^t\Tr\stability\,dt'}
= 1.
\end{equation}
The norm-preserving evolution operator is then orthogonal: $\tilde{\jacobian}\tilde{\jacobian}^\top = \mathbb{I}$ and $\tilde{\jacobian}^\top = \tilde{\jacobian}^{-1}$.
The similarity transform is then also an orthogonal transformation.
Another way to see this is: $|\jacobian|=\prod_i^ne^{\int_{t_0}^t\lambda_idt'}=e^{\int_{t_0}^t\sum_i\lambda_idt'}$.

\section{Basis states for density matrix representation}\label{sec:basis_rep}

\noindent Consider a pure state formed by the basis vector $\ket{\delta\bm{\phi}_i}\in T\mathcal{M}$ 
\begin{align}
\brho_i(t) & = \ket{\delta\bm{\phi}_i(t)}\bra{\delta\bm{\phi}_i(t)}.
\end{align}
There are uncountably many sets of linearly independent vectors that span an $n$-dimensional phase space.
A simple choice is the Cartesian coordinates fixed on the given trajectory, $\ket{\delta \bm{\phi}_i} = \delta_{ij}\ket{\mathbb{1}}$, where $\delta_{ij}$ is the Kronecker delta, $i=1,\ldots,n$, and $\ket{\mathbb{1}}$ represents an $n\times 1$ unit matrix.
The instantaneous Lyapunov exponents for these basis states are the diagonal elements of the stability matrix $\stability$:
\begin{align}
r_i = \bra{\delta \bm{\phi}_i}\stability\ket{\delta \bm{\phi}_i} = \Tr(\stability_+\brho_i) = (\boldsymbol{A})_{ii}.
\end{align}
Other coordinates systems can be chosen to represent the basis states that are time-independent or time-dependent and comoving with the phase point.

\smallskip
\noindent\textit{a. $\stability_{+}$ basis.--} Eigenvalues of $\stability_{+}$, $\sigma^{\stability_+}_{i}$, satisfy
$\stability_{+}\brho_i = \sigma^{\stability_+}_{i}\brho_i$.
The instantaneous Lyapunov exponents are tangent space averages:
\begin{align}
r_i^{\stability_+} &= \Tr(\stability_+\brho_i) = \sigma^{\stability_+}_{i}.
\end{align}
Due to the symmetric nature of $\stability_+$, its eigenvalues are always real.
Also, notice that in the equation of motion for $\brho_i$ (Eq.~\ref{equ:EOM_rho}), the first and the third terms cancel out and the equation simplifies:
\begin{equation}
  \frac{d\brho_i}{dt} = [\stability_-,\brho_i].
\end{equation}
This basis spans the $n$-dimensional phase space, so the equation of motion for the density matrix $\brho$
leads to the generalized Liouville equation,
\begin{equation}
 \frac{d|\brho|}{dt} = \left|\sum_{i=1}^n[\stability_-,\brho_i]\right|
  = \bigl\rvert[\stability_-,\brho]\bigr\rvert = 0,
\end{equation}
and the time invariance of the normalized density matrix.

\smallskip
\noindent\textit{b. $\stability_-$ basis.--}
The eigenvectors of the anti-symmetric matrix $\stability_-$ form a complete basis but are not necessarily orthogonal.
The eigenvalues of $\stability_-$ are purely imaginary, so: $[\stability_-,\brho_i] = \stability_-\brho_i - \brho_i\stability_- = \mathbb{0}.$ Eq.~\ref{equ:EOM_rho} then becomes
\begin{equation}
\frac{d\brho_i}{dt} = \{\stability_+,\brho_i\} -  2r_i^{\stability_-}\brho_i,
\end{equation}
with the instantaneous Lyapunov exponent $r_i^{\stability_-}$ in the eigenbasis of $\stability_-$. The generalized Liouville equation for the maximally mixed state $\brho$ in this basis is then:
\begin{equation}
  \frac{d|\brho|}{dt} = \bigl\rvert\{\stability_+,\brho\} - 2r\brho\bigr\rvert = 0.
\end{equation}

\smallskip
\noindent\textit{c. $\stability$ basis.--} The eigenvectors of $\stability$ form a complete set of basis but not mutually orthogonal. If the eigenvalues of $\stability$ are denoted by $\sigma^{\stability}_i$, then the instantaneous Lyapunov exponents for these basis states are derived as follows
\begin{align}
r_i^{\stability} &= \Tr(\stability_+\brho_i) = \frac{1}{2}\Tr(\stability\brho_i + \stability^\top\brho_i) \nonumber\\
&= \frac{1}{2}(\sigma^{\stability}_i + \sigma_i^{\dagger\stability}) =\Re(\sigma^{\stability}_i)\label{eq:ILE_A},
\end{align}
where $\dagger$ denotes the complex conjugate and $\Re(.)$ gives the real part. We use this expression of $r_i^A$ to write Eq.~\ref{equ:EOM_rho} in the $\stability$ basis:
\begin{align}
	\frac{d\brho_i}{dt} &= \stability\brho_i + \brho_i\stability^\top - 2r^A_i\brho_i \nonumber\\
	&= \sigma^A_i\brho_i+\sigma_i^{\dagger\stability}\brho_i -2r^A_i\brho_i, \nonumber\\
	&=2\Re(\sigma_i^\stability)\brho_i-2r^A_i\brho_i = \mathbb{0}.
\end{align}
In this basis, the maximally mixed state $\brho$ is a constant of motion and the generalized Liouville equation becomes $|\dot{\boldsymbol{\rho}}|=0$. 

\section{Density matrix formulation for Hamiltonian systems}\label{sec:DMD_Hamiltonian}

For Hamiltonian systems, the components of a Lyapunov vector are the first variations of
position and momentum $(\delta \boldsymbol{q}, \delta \boldsymbol{p})$ and the stability matrix $\stability$ becomes:
\begin{eqnarray}\nonumber
\stabilityH&=&\left(
\begin{array}{cc}
\partial_\bmQ\partial_\bmP\Hamiltonian & \partial_\bmP^2\Hamiltonian\\
-\partial_\bmQ^2\Hamiltonian & -\partial_\bmP\partial_\bmQ^2\Hamiltonian\\
\end{array}\right)=\left(
\begin{array}{cc}
\mathbb{0} & \partial_\bmP^2\Hamiltonian\\
-\partial_\bmQ^2\Hamiltonian & \mathbb{0}\\
\end{array}\right).
\end{eqnarray}
The trace of the Jacobian gives the traditional form of Liouville's theorem.
In the last equality, we assume the Hamiltonian is $H(\boldsymbol{q},\boldsymbol{p}) = T(\boldsymbol{p}) + V(\boldsymbol{q})$ and the stability matrix is the product of its Hessian and the Poisson matrix:
\begin{equation*}
\bm{\Omega}=\left(
\begin{array}{cc}
\mathbb{0} & \mathbb{I}\\
-\mathbb{I} & \mathbb{0}\\
\end{array}\right).
\end{equation*}
The identity matrices $\mathbb{I}$ are $n/2\times n/2$.

The symmetric and anti-symmetric parts of the stability matrix are:
\begin{align}
\stabilityH_+ &= \frac{1}{2}(\partial_\bmP^2\Hamiltonian-\partial_\bmQ^2\Hamiltonian)\left(
\begin{array}{cc}
\mathbb{0} & \mathbb{I}\\
\mathbb{I} & \mathbb{0}\\
\end{array}\right) \\
\stabilityH_- &= \frac{1}{2}(\partial_\bmP^2\Hamiltonian+\partial_\bmQ^2\Hamiltonian)\left(
\begin{array}{cc}
\mathbb{0} & \mathbb{I}\\
-\mathbb{I} & \mathbb{0}\\
\end{array}\right),
\end{align}
respectively. Equation~\ref{equ:EOM_rho} then becomes:
\begin{align}
\frac{d\brho}{dt} &= \{\stabilityH_+,\brho\} + [\stabilityH_-,\brho] - 2r\brho.
\label{Liouville_Lyap_eq_H_dynamics}
\end{align}
Since the traditional forms for Liouville's theorem/equation are specific to Hamiltonian dynamics, we consider eigenbases described in Appendix~\ref{sec:basis_rep} for this special case.

\smallskip
\noindent\textit{a. $\stabilityH_{+}$ basis.--} 
The eigenvectors of $\stabilityH_{+}$ are a complete set of orthonormal basis vectors in an $n$-dimensinal phase space, given in a general form by
\begin{align}
	\ket{\delta \bphi_{qj}} = \frac{1}{\sqrt{2}}(\delta_{jk}\ket{\mathbb{1}}+\delta_{j'k'}\ket{\mathbb{1}}),\nonumber \\
	\ket{\delta \bphi_{pj}} = \frac{1}{\sqrt{2}}(\delta_{jk}\ket{\mathbb{1}}-\delta_{j'k'}\ket{\mathbb{1}}),
\end{align}
where $\delta_{jk}$ and $\delta_{j'k'}$ are Kronecker deltas: $\delta_{jk} = 1$ for $j=k$ and zero otherwise. Indices $j'$ and $j$ are related: $j' = n/2+j$, where $j$ runs from $1$ to $n/2$.

For example, in the case of a 2D-Hamiltonian system, these basis vectors become
\begin{align}
\ket{\delta \bm{\phi}_1} &= \frac{1}{\sqrt{2}}\begin{pmatrix}
1\\ 
1
\end{pmatrix},\quad\ket{\delta\bm{\phi}_2} =\frac{1}{\sqrt{2}}\begin{pmatrix}
1\\ 
-1\\
\end{pmatrix}
\end{align} 
that define the basis states:
\begin{align}
\brho_1 &=  \frac{1}{2}
\begin{pmatrix}
1 &	1\\ 
1 &	1
\end{pmatrix},\quad
\brho_2 =  \frac{1}{2}\begin{pmatrix}
1 &	-1\\ 
-1 & 1\\
\end{pmatrix}.
\end{align}
For these states, the instantaneous Lyapunov exponents are the eigenvalues of $\stabilityH_{+}$,
\begin{align}\label{eq:ILE_symA}
r_k^{\stabilityH_+} = \pm  \frac{1}{2}(\partial_\bmP^2\Hamiltonian-\partial_\bmQ^2\Hamiltonian)=  \pm  \frac{1}{2}\,\grad\cdot 
\begin{pmatrix} \dot p \\ \dot q \end{pmatrix}.
\end{align}

\smallskip
\noindent\textit{b. $\stabilityH_{-}$ basis.--} The eigenvectors of $\stabilityH_-$ are also a complete set of orthonormal basis vectors,
\begin{align}
\ket{\delta \bphi_{qj}} = \frac{1}{\sqrt{2}}(\delta_{jk}\ket{\mathbb{1}}+i\delta_{j'k'}\ket{\mathbb{1}}),\nonumber \\
\ket{\delta \bphi_{pj}} = \frac{1}{\sqrt{2}}(\delta_{jk}\ket{\mathbb{1}}-i\delta_{j'k'}\ket{\mathbb{1}}),
\end{align}
where $\delta_{jk}$ and $\delta_{j'k'}$ are Kronecker deltas with $j' = n/2+j$.

Again, for a 2D-Hamiltonian system, these states become
\begin{align}
\ket{\delta \bphi_1} &= \frac{1}{\sqrt{2}}\begin{pmatrix}
1\\ 
i
\end{pmatrix}, \quad
\ket{\delta \bphi_2}=\frac{1}{\sqrt{2}}\begin{pmatrix}
1\\ 
-i
\end{pmatrix},
\end{align}
with purely imaginary eigenvalues $\pm i(\partial_\bmP^2\Hamiltonian+\partial_\bmQ^2\Hamiltonian) $.
The corresponding basis states are:
\begin{align}\label{eq:basis_states_A_anti}
\brho_1 &=  \frac{1}{2}
\begin{pmatrix}
1 & -i\\ 
i &1
\end{pmatrix},\quad
\brho_2 =  \frac{1}{2}\begin{pmatrix}
1 &	i\\ 
-i & 1\\
\end{pmatrix}.
\end{align}
For these basis states,
\begin{align}
  r_k^{\stability_-} & = \Tr(\stabilityH_+\brho_k) = 0,
\end{align} 
the instantaneous Lyapunov exponents in the $\stabilityH_-$ basis always vanishes for Hamiltonian systems.
This fact further simplifies Eq.~\ref{equ:EOM_rho} to,
\begin{equation}
  \frac{d\brho_k}{dt} = \{\stability_+,\brho_k\},
\end{equation}
and equation of motion that depends only on the classical commutator.

\smallskip
\noindent\textit{c. $\stabilityH$ basis.--} From Eq.~\ref{eq:ILE_A}, we know that the instantaneous Lyapunov exponents in this basis are the real part of the eigenvalues of $\stabilityH$,
\begin{align}
\lambda^{\stabilityH}_k &= \pm i \sqrt{\partial^2_{p_k}\Hamiltonian\,\partial^2_{q_k} H}.
\end{align}
So, for non-zero exponents, we must have $\partial^2_{p_k}\Hamiltonian\,\partial^2_{q_k} H<0$. Equation~\ref{Liouville_Lyap_eq_H_dynamics} in this basis becomes $d_t\brho_k = \mathbb{0}$.

\smallskip
\noindent\textit{d. Tangent pure states.--} Consider conjugate pure states in the tangent space for Hamiltonian dynamics, as introduced in the main text. We write them for a 2D-Hamiltonian system in Eq.~\ref{eq:tangent_pure_states}:
\begin{align}
\|\dot \ex\|^2\brho_{\dot \ex} &= 
\begin{pmatrix}
(\partial_p \Hamiltonian)^2 & -\partial_p \Hamiltonian\partial_q \Hamiltonian\\[2pt]
-\partial_q \Hamiltonian\partial_p \Hamiltonian & (\partial_q \Hamiltonian)^2\\ 
\end{pmatrix}=
\begin{pmatrix}
\dot q^2 & \dot p\dot q \\[2pt]
\dot p\dot q & \dot p^2\\ 
\end{pmatrix}, \nonumber 
\end{align}
\begin{align}
\|\grad\Hamiltonian\|^2\brho_{\grad\Hamiltonian} &= 
\begin{pmatrix}
(\partial_q \Hamiltonian)^2 & \partial_q \Hamiltonian\partial_p \Hamiltonian\\[2pt]
\partial_p H\partial_q \Hamiltonian & (\partial_p \Hamiltonian)^2\\ 
\end{pmatrix}=
\begin{pmatrix}
\dot p^2 & -\dot p\dot q\\
-\dot p\dot q & \dot q^2
\end{pmatrix}. \nonumber
\end{align}
For an $n-$dimensional system, these pure states are formed by the following basis vectors:
\begin{align}
\ket{\delta \bphi_{\dot\ex_j}} &= \frac{1}{\|\dot{\ex_j}\|}(\dot q_j\delta_{jk}\ket{\mathbb{1}}+\dot p_j\delta_{j'k'}\ket{\mathbb{1}}),\nonumber \\
\ket{\delta \bphi_{\nabla H_j}} &=\frac{1}{\|\nabla H_j\|}(-\dot p_j\delta_{jk}\ket{\mathbb{1}}+\dot q_j\delta_{j'k'}\ket{\mathbb{1}}),
\end{align}
with Kronecker deltas $\delta_{jk}$ and $\delta_{j'k'}$ and $j' = n/2+j$.

For these time-dependent basis states, the instantaneous Lyapunov exponents are:
\begin{align}\label{eq:ILE_tangent}
r_k &= \pm  \frac{\dot q_k\dot p_k}{||\dot \ex_k||^2}\grad\cdot 
\begin{pmatrix} \dot p_k \\ \dot q_k \end{pmatrix} = \pm \Tr(\brho_{\dot \ex_k}\stabilityH_+).
\end{align}
Table~\ref{table:summary} (main text) shows the summary of results in these bases for Hamiltonian and dissipative systems.

\section{Case studies}\label{sec:case_examples}
\setcounter{figure}{0}
\makeatletter 
\renewcommand{\thefigure}{F\arabic{figure}}
\noindent\textit{a. The linear harmonic oscillator.--}
\vspace*{-0.5cm}
\begin{center}
\begin{figure}[h!]
	\includegraphics[width=4.12cm]{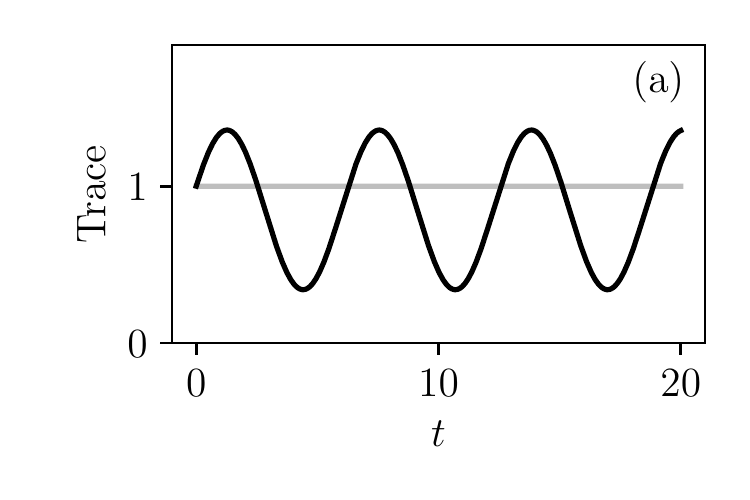}\\
	\includegraphics[width=4.12cm]{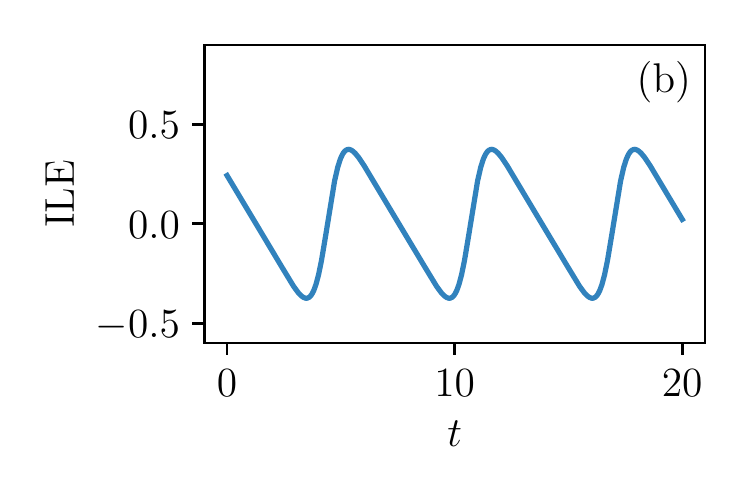} \includegraphics[width=4.2cm]{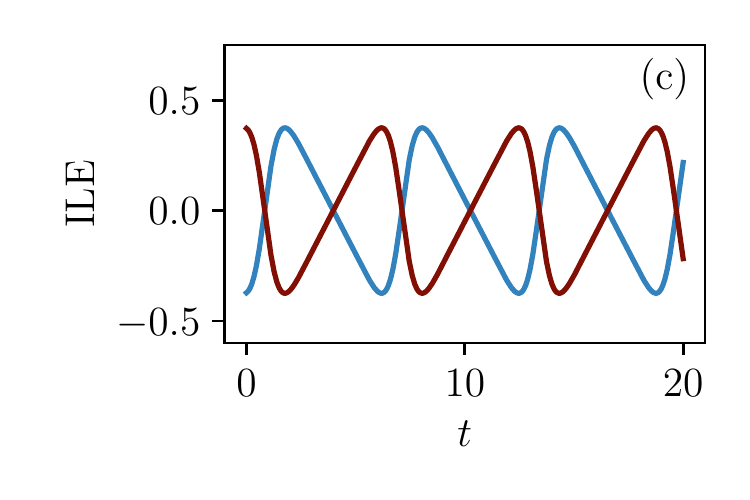}\\
	\caption{\label{fig:SI_HO} (a) The linear harmonic oscillator ($\omega=0.5$): $\Tr\bxi$ (black) and $\Tr\brho$ (gray) as a function of time, (b) Time evolution of instantaneous Lyapunov exponent (ILE) for an arbitrary pure perturbation state, and (c) ILEs in the conjugate tangent space directions.}
\end{figure}
\end{center}

\noindent\textit{b. The damped harmonic oscillator.--}
\begin{center}
	\begin{figure}[h!]
		\vspace*{-0.5cm}
		\includegraphics[width=4.12cm]{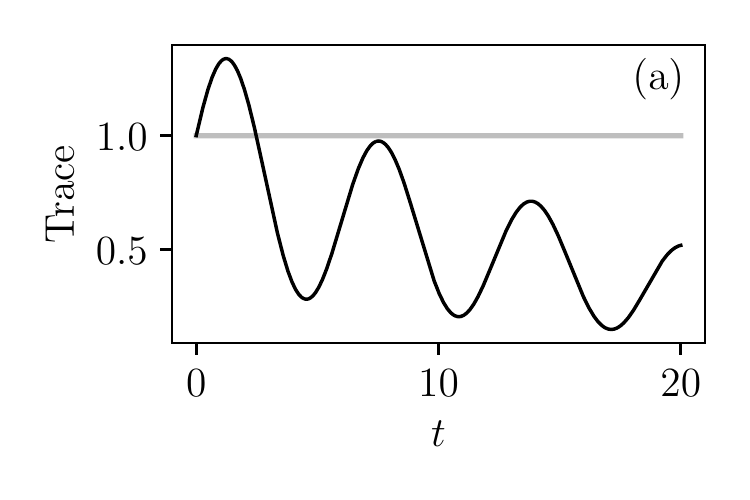}\\
		\includegraphics[width=4.12cm]{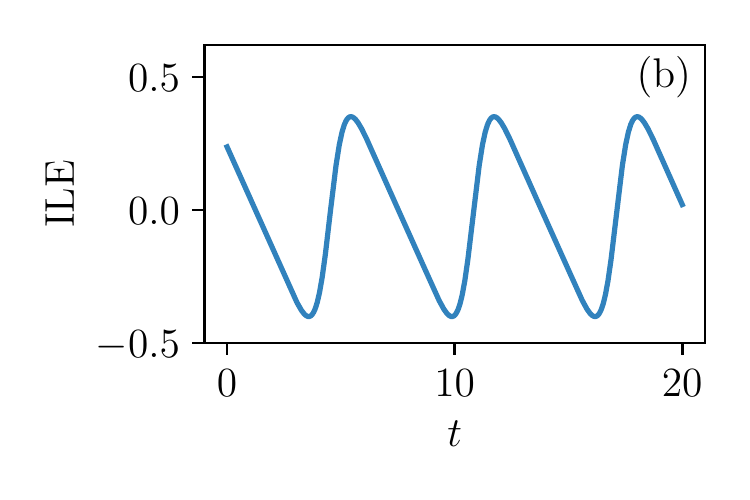} \includegraphics[width=4.2cm]{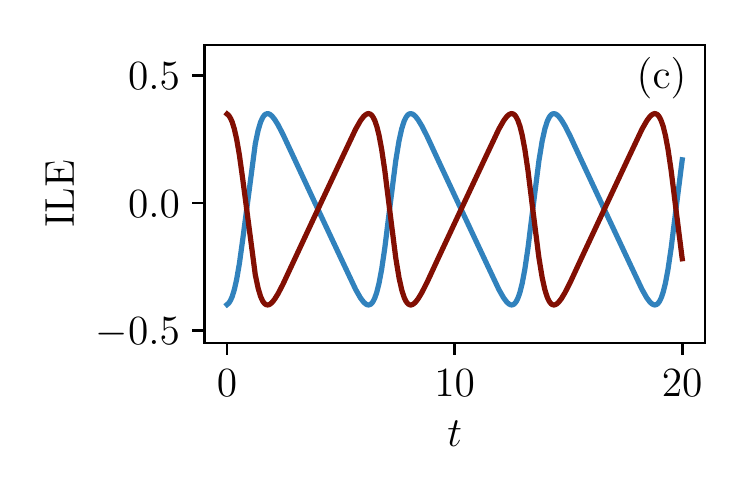}
		\caption{\label{fig:SI_DHO} Damped harmonic oscillator ($\omega=0.5$, $\gamma=0.05$): (a) Time evolution of $\Tr\bxi$ (black) and $\Tr\brho$ (gray) and (b) instantaneous Lyapunov exponent (ILE) for an arbitrary pure perturbation state, and (c) ILEs in the conjugate tangent space directions.}
	\end{figure}
\end{center}

\noindent\textit{c. The H\'enon-Heiles system.--}

The equations of motion,
\begin{align}
\dot{x} &= p_x, \quad \dot{y} = p_y,\nonumber\\
\dot{p_x} &= -x - 2xy, \quad \dot{p_y} = -y - x^2 + y^2,
\end{align}
lead to the stability matrix:
\begin{align}
\stabilityH =
\begin{pmatrix}
0 & 0 & 1 & 0\\
0 & 0 & 0 & 1\\
-1 - 2y & -2x & 0 & 0\\
-2x & -1+2y & 0 & 0
\end{pmatrix}.
\end{align}
\begin{figure}[ht!]
	\includegraphics[width=4.25cm]{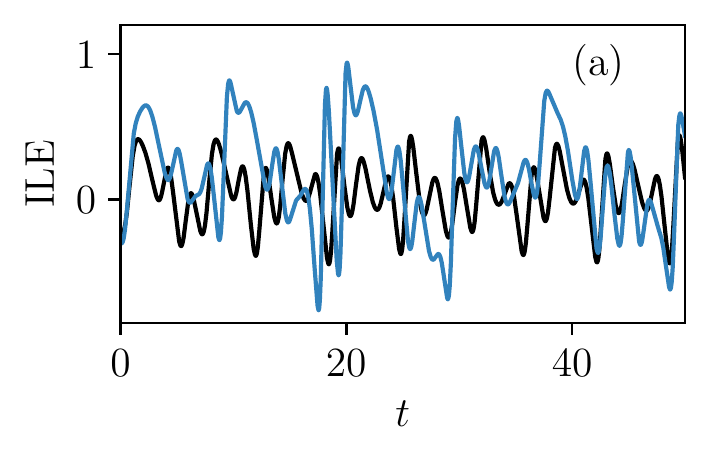}	\includegraphics[width=4.25cm]{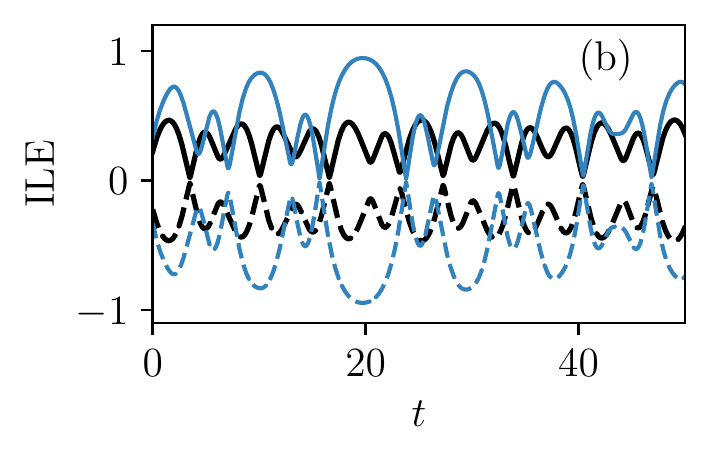}\\
	\caption{\label{fig:SI_HH} The H\'enon-Heiles model: for a regular orbit with energy $E=0.0833$ and a chaotic orbit with energy $E=0.1677$ (a) instantaneous Lyapunov exponent (ILE) for arbitrary pure states. (b) ILEs in the $\stabilityH_+$ basis (black solid and dashed lines for the regular orbit and blue solid and dashed lines for the chaotic orbit).}
\end{figure}

Figure~\ref{fig:SI_HH}(a) shows the time evolution of a pure state described by an arbitrary unit vector on a regular and a chaotic orbit. The orbits chosen corresponds to $E=0.0833$ (regular) and $E=0.1667$ (chaotic).\\

Symmetric and anti-symmetric parts of the stability matrix are given by
\begin{align}
\stabilityH_+ &=
\begin{pmatrix}
0 & 0 & -y & -x\\
0 & 0 & -x & y\\
-y & -x & 0 & 0\\
-x & y & 0 & 0
\end{pmatrix},
\end{align}
In the eigenbasis of $\stabilityH_+$, the instantaneous Lyapunov exponents are, as usual, the eigenvalues
$r_{1,2} = \pm 2y$, and $r_{3,4} = \mp 2y$.
Figure~\ref{fig:SI_HH}(b) shows their time evolution.
The instantaneous Lyapunov exponents in the tangent space directions $\dot\ex$ and $\grad H$ are:
\begin{align}
r_{1,2} & = \pm \frac{2y(x+2xy)p_x}{p_x^2+(x+2xy)^2},
\end{align}
\begin{align}
r_{3,4} & = \mp \frac{2y(x^2-y^2+y)p_y}{p_y^2+(x^2-y^2+y)^2}.
\end{align}

\vspace*{0.5cm}

\noindent\textit{b. The Lorenz system.--}
As another example, we consider the model of atmospheric convection due to Lorenz and Fetter~\cite{Lorenz63}.
The model is defined by the ordinary differential equations,
\begin{align}
\dot{x} &= \sigma(y-x), \quad\dot{y} = x(\rho - z) -y, \quad \dot{z} = xy - \beta z,
\end{align}
\begin{figure}[t]
	\includegraphics[width=4.5cm]{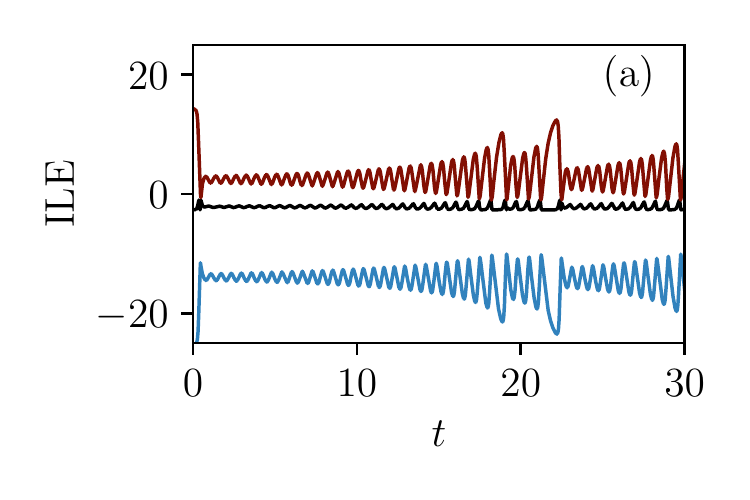}\includegraphics[width=4.5cm]{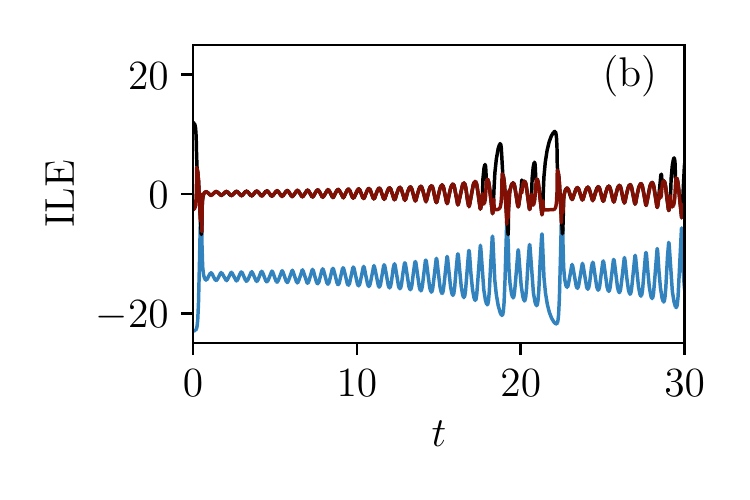}\\
	\caption{\label{fig:SI_Lorenz} The Lorenz model ($\sigma = 10$, $\beta = 8/3$, $\rho = 28$): Time evolution of instantaneous Lyapunov exponents (ILEs) in (a) the $\stability_+$ basis and (b) the $\stability$ basis.}
\end{figure}
with the stability matrix:
\begin{align}
\stability &=\begin{pmatrix}
-\sigma & \sigma & 0\\
\rho-z & -1 & -x\\
y & x & -\beta
\end{pmatrix}.
\end{align}
The instantaneous Lyapunov exponents can be computed in any of the eigenbases using the symmetric and part of the stability matrix:
\begin{align}
\stability_+&=\begin{pmatrix}
-\sigma & \frac{1}{2}(\sigma+\rho-z) & \frac{1}{2}y\\
\frac{1}{2}(\sigma+\rho-z) & -1 & 0\\
\frac{1}{2}y & 0 & -\beta
\end{pmatrix}.
\end{align}
For the $\stability_+$ and $\stability$ bases, they are shown in Figs.~\ref{fig:SI_Lorenz}(a) and ~\ref{fig:SI_Lorenz}(b) respectively, for  $\sigma = 10$, $\beta = 8/3$, $\rho = 28$.


%

\end{document}